\documentclass[a4paper,11pt]{article}
\pdfoutput=1

\usepackage{jheppub1}
\usepackage[T1]{fontenc}
\usepackage{amssymb}
\usepackage{graphicx}
\usepackage{amsmath}
\usepackage{tensor}
\usepackage{hyperref}
\usepackage{epstopdf}
\usepackage{extarrows}
\usepackage{xcolor}
\usepackage{tensor}
\usepackage{subfigure}

\newcommand{\td}{\text{d}}

\def\I {\hat{\mathbb{I}}}
\def\x {\hat{x}}

\def\O {\hat{O}}
\def\U {\hat{U}}
\def\C {\mathcal{C}}
\def\F {\tilde{F}}

\def\Tr {\text{Tr}}

\newcommand{\kyr}[1]{{\color{red}{#1}}}

\begin{document}
\title{Time evolution of the complexity in chaotic systems: a concrete example}
\author[a,b]{Run-Qiu Yang}
\author[c]{and Keun-Young Kim}

\emailAdd{aqiu@tju.edu.cn}
\emailAdd{fortoe@gist.ac.kr}

\affiliation[a]{Center for Joint Quantum Studies and Department of Physics School of Science, Tianjin University, Yaguan Road 135, Jinnan District, 300350 Tianjin, P.~R.~China}
\affiliation[b]{Quantum Universe Center, Korea Institute for Advanced Study, Seoul 130-722, Korea}
\affiliation[c]{ School of Physics and Chemistry, Gwangju Institute of Science and Technology,
Gwangju 61005, Korea
}

\abstract{
We investigate the time evolution of the complexity of the operator by the Sachdev-Ye-Kitaev (SYK) model with $N$ Majorana fermions. We follow Nielsen's idea of complexity geometry and geodesics thereof. We show that it is possible that the bi-invariant complexity geometry can exhibit the conjectured time evolution of the complexity in chaotic systems: i) linear growth until $t\sim e^{N}$, ii) saturation and small fluctuations after then.
We also show that the Lloyd's bound is realized in this model.  Interestingly, these characteristic features appear only if the complexity geometry is the most natural ``non-Riemannian'' Finsler geometry. This serves as a concrete example showing that the bi-invariant complexity may be a competitive candidate for the complexity in quantum mechanics/field theory (QM/QFT). We provide another argument showing a naturalness of bi-invariant complexity in QM/QFT.
That is that the bi-invariance naturally implies the equivalence of the right-invariant complexity and left-invariant complexity, either of which may correspond to the complexity of a given operator.  Without bi-invariance, one needs to answer why only right (left) invariant complexity corresponds to the ``complexity'', instead of only left (right) invariant complexity.

}
\maketitle

\noindent

\section{Introduction}
The concepts of the quantum information theory have been widely applied to the studies of quantum field theories and gravity. In particular, the  quantum ``complexity'', a concept originated from the quantum circuits and quantum computations, has been introduced to the studies of black hole physics~\cite{Harlow:2013tf} and the AdS/CFT correspondence~\cite{Susskind:2014rva,Susskind:2014rva2,Stanford:2014jda}. Roughly speaking, from the perspective of quantum circuit,  the complexity of an operator (circuit) is the minimal number of ``elementary operations" (gates) to construct the operator (circuit).

In order to compute the complexity by using gravity theory in the context of the AdS/CFT correspondence, Refs.~\cite{Stanford:2014jda} and \cite{Brown:2015bva} proposed two holographic conjectures called the complexity-volume (CV) conjecture and the complexity-action (CA) conjecture. The CV conjecture states  that  the complexity of a boundary state is given by the  maximum volume of all spacelike surfaces at fixed time slices at the AdS boundary. The CA conjecture states that the complexity of a boundary state is given by the on-shell action in Wheeler-DeWitt (WdW) patch.
After these two holographic conjectures were proposed, many works have been done to study the properties of the CA and CV conjectures, such as the upper bound of the complexity growth rate in various gravity systems~\cite{Cai:2016xho,Yang:2016awy,Pan:2016ecg,Alishahiha:2017hwg,An:2018xhv,Jiang:2018pfk,Jiang:2018sqj}, the UV divergent structures of the CV and CA conectures~\cite{Chapman:2016hwi,Kim:2017lrw}, the time-evolution of the complexity in CV and CA conjectures~\cite{Carmi:2017jqz,Kim:2017qrq,An:2018dbz}, the quench effects in the holographic complexity~\cite{Moosa:2017yvt,Chen:2018mcc} and so on. Besides these two conjectures, other conjectures for the complexity, such as sub-region complexity, were also proposed in holography for different systems and purposes (see, for example, Refs~\cite{Alishahiha:2015rta,Ben-Ami:2016qex,Couch:2016exn,Caputa:2017urj,Caputa:2017yrh}).



Though much progress on the complexity in gravity side has been made, the precise meaning and a well-proposed definition of the complexity in field theory side is still incomplete. The first attempt to find a generalization of the {\it discrete} quantum  circuit complexity to {\it continuous} systems was proposed by Nielsen et al.~\cite{Nielsen1133,Nielsen:2006:GAQ:2011686.2011688,Dowling:2008:GQC:2016985.2016986}. The basic idea is to identify complexity with the geodesic distance in a certain geometry called  ``complexity geometry''.
Built on this idea for quantum {\it circuit}, there have been many attempts to define the complexity in {\it quantum field theory}  (QFT)~\cite{Brown:2017jil,Jefferson:2017sdb,Yang:2017nfn,Chapman:2017rqy,Khan:2018rzm,Camargo:2018eof,Chapman:2018hou,Susskind:2014jwa,Brown:2016wib,Hashimoto:2017fga,Hashimoto:2018bmb,Flory:2018akz,Flory:2019kah,Belin:2018fxe,Belin:2018bpg,Yang:2018nda,Yang:2018tpo,Yang:2018cgx,Balasubramanian:2019wgd,Yang:2019udi}\footnote{There are also similar but a little different proposals, such as  the Fubini-study metric~\cite{Chapman:2017rqy} and path-integral optimization~\cite{Bhattacharyya:2018wym,Takayanagi:2018pml}.}.
Thus, the problem boils down to how to construct the ``complexity geometry'' and investigate its geometric properties.

In many works, the complexity geometry is {\it assumed} to be only right-invariant but not left-invariant. i.e. the complexity geometry is assume to be non-bi-invariant (bi-invariant means both right-invariant and left-invariant). The meanings of these invariances will be explained in detail around Eq.~\eqref{ppp1}.
One of the reasons why non-bi-invariant complexity is {\it assumed} is that it is often {\it claimed}~\cite{Brown:2017jil,Balasubramanian:2019wgd} that a bi-invariant  complexity cannot reproduce the ``expected'' time evolution of the complexity in chaotic systems. The expected time evolution is as follows. For a chaotic system with $N$ degrees of freedom, the complexity evolves as time goes in three stages: i) linear growth until $t\sim e^{N}$; ii) saturation and small fluctuations after then; iii) and quantum recurrence at $t\sim e^{e^N}$. However, this {\it claimed }incompatibility between ``expected'' time evolution and bi-invariance has not been checked until the recent paper Ref.~\cite{Balasubramanian:2019wgd} appeared. In this very interesting work, the authors claimed that the aforementioned incompatibility was shown in a concrete mode, the Sachdev-Ye-Kitaev (SYK) model.

One of the goals of this paper is to revisit the model studied in Ref.~\cite{Balasubramanian:2019wgd} and show that it is indeed {\it possible} that bi-invariant complexity is compatible to the ``expected'' time evolution. We stress that we do not claim that the analysis in Ref.~\cite{Balasubramanian:2019wgd} is not correct. We want to point out that it is still possible to have bi-invariant complexity compatible with the ``expected'' time evolution in the SYK model. The key point of this possibility is the fact that the {\it complexity growth rate} is dimensionful so it depends on the time scale, while the complexity itself does not.
Furthermore, with this bi-invariant complexity, we also obtain  more interesting results i) the Lloyd's bound can be realized; ii) for complexity geometry, non-Riemannian Finsler geometry is favored than Riemannian geometry.

Based on our analysis on a chaotic system, it seems that a bi-invariant complexity is still a viable and competitive one in the complexity theory compared with an only-right-invariant one. Indeed, we have provided various arguments to support bi-invariant complexity in our series of works~\cite{Yang:2018nda,Yang:2018tpo,Yang:2018cgx,Yang:2019udi}. In this paper,
we want to provide yet another argument to support bi-invariant complexity, which is another goal of this paper. It is inspired by a simple question: it is mathematical fact that we may define both only-right-invariant complexity and only-left-invariant complexity for a given operator. Are these two equivalent? If not, which one corresponds to {\it the} ``complexity''? Naively, one may think that i) one can choose either an only-right-invariant (not bi-invariant) complexity or an only-left-invariant (not bi-invariant) complexity ii) it is possible that these two choices give us equivalent physics. However, we will show that if they are not bi-invariant then they cannot give equivalent physics.

This paper is organized as follows. In section~\ref{csyk},  we first make a short overview on the complexity of operator and explain there are two different ways (right-invariant and left-invariant way) in defining complexity.  In section~\ref{sec3}, as a concrete example, we consider the SYK model to show how to compute the complexity growth. By using both analytic and numerical methods we demonstrate that bi-invariant complexity indeed grows linearly up to an exponential time scale. Here, we also show that the Lloyd's bound can be realized and the complexity geometry is non-Riemannian Finsler geometry.  In section~\ref{qonr} we explain a consequence if the complexity is not bi-invariant: there are two ways (left or right-invariant ways) in defining complexity but these two will not be equivalent.

\section{Review on complexity of operators}\label{csyk}
Let us first make a brief overview on how to define the complexity for operators, which has been explained in Refs.~\cite{Yang:2018nda,Yang:2018tpo}.

\paragraph{General ansatz for complexity}

We denote a complexity of an operator $\hat{x}$ in a finite dimensional special unitary group SU($n$) by $\mathcal{C}(\hat{x})$.
Mathematically, the complexity $\C$  is a map from SU($n$) to $\mathbb{R}^+\cup\{0\}$ which satisfies the following four axioms,
\begin{enumerate}
\item[\textbf{G1}] (\textit{Nonnegativity})\\
$\forall \x\in$ {SU($n$)}, $\mathcal{C}(\x)\geq0$, and $\C(\x)=0$ if $\x $ is the identity
\vspace{-0.2cm}
\item[\textbf{G2}] (\textit{Series decomposition rule})\\
$\forall \x,\hat{y}\in$ {SU($n$)}, $\mathcal{C}(\x)+\mathcal{C}(\hat{y})\geq\mathcal{C}(\x\hat{y})$
\vspace{-0.2cm}
\item[\textbf{G3}]  ({\textit{Parallel decomposition rule}}) \\
$\forall  \x_1 ,\x_2$, in a matrix representation, $\C(\x_1\oplus\x_2)^p=\C(\x_1)^p+\C(\x_2)^p$ with a nonnegative number $p$\,
\vspace{-0.2cm}
\item[\textbf{G4}] (\textit{Smoothness})
For any infinitesimal operator in SU($n$), $\delta \hat{O} = \exp (iH\delta s)$, the complexity is a smooth function of  $H \ne 0$ and $\delta s \ge 0$, i.e.,
\begin{equation} \label{compde0}
\mathcal{C} (\delta \hat{O}) =\tilde{F}(H) \delta s + \mathcal{O} (\delta s^2) \,
\end{equation}
\end{enumerate}%
Here, \textbf{G1} and \textbf{G2} are natural concepts coming from the notion of ``distance'' and \textbf{G2} corresponds to the triangle inequality.  In \textbf{G2}  and \textbf{G3}, we consider the situation where an operator can be decomposed in part. From the quantum circuit perspective \textbf{G2} is for serial decomposition and \textbf{G3} is for parallel decomposition of a bigger circuit.
Based on the intuition of the complexity of quantum circuits we make two ansatzes, \textbf{G2}  and \textbf{G3}, which give the relation between the  complexity in total and the complexity in part.

As we mentioned, \textbf{G2} is kind of a triangle inequality but \textbf{G3} may have a freedom of $p$ in principle.
From physical viewpoint, the most natural choice is $p=1$. Intuitively, it is because the complexity of the circuit composed of two independent parallel sub-circuits will be the sum of the complexity of sub-circuits:
\begin{equation}
\C(\x_1\oplus\x_2)=\C(\x_1)+\C(\x_2) \,.
\end{equation}
From mathematical viewpoint, the choice of $p=2$
\begin{equation}
\C(\x_1\oplus\x_2)^2=\C(\x_1)^2+\C(\x_2)^2 \,,
\end{equation}
may be convenient and {\it familiar} because it turns out that the complexity geometry becomes Riemannian, while for $p \ne 2$ the complexity geometry is non-Riemannian Finsler geometry.\footnote{For more details on \textbf{G3}, we refer to Fig.2 and section 2.2 in \cite{Yang:2018nda} or section 2 in~\cite{Yang:2018tpo}.}
In addition to a good physical intuition for $p=1$, based on the properties of the complexity of the SYK model in next section, we will provide another supporting evidence that the $p=1$ case is indeed physically favored than the widely discussed $p=2$ case of the Riemannian geometry or other $p\ne1$ cases.

\begin{figure}
 \centering
  \includegraphics[width=.4\textwidth]{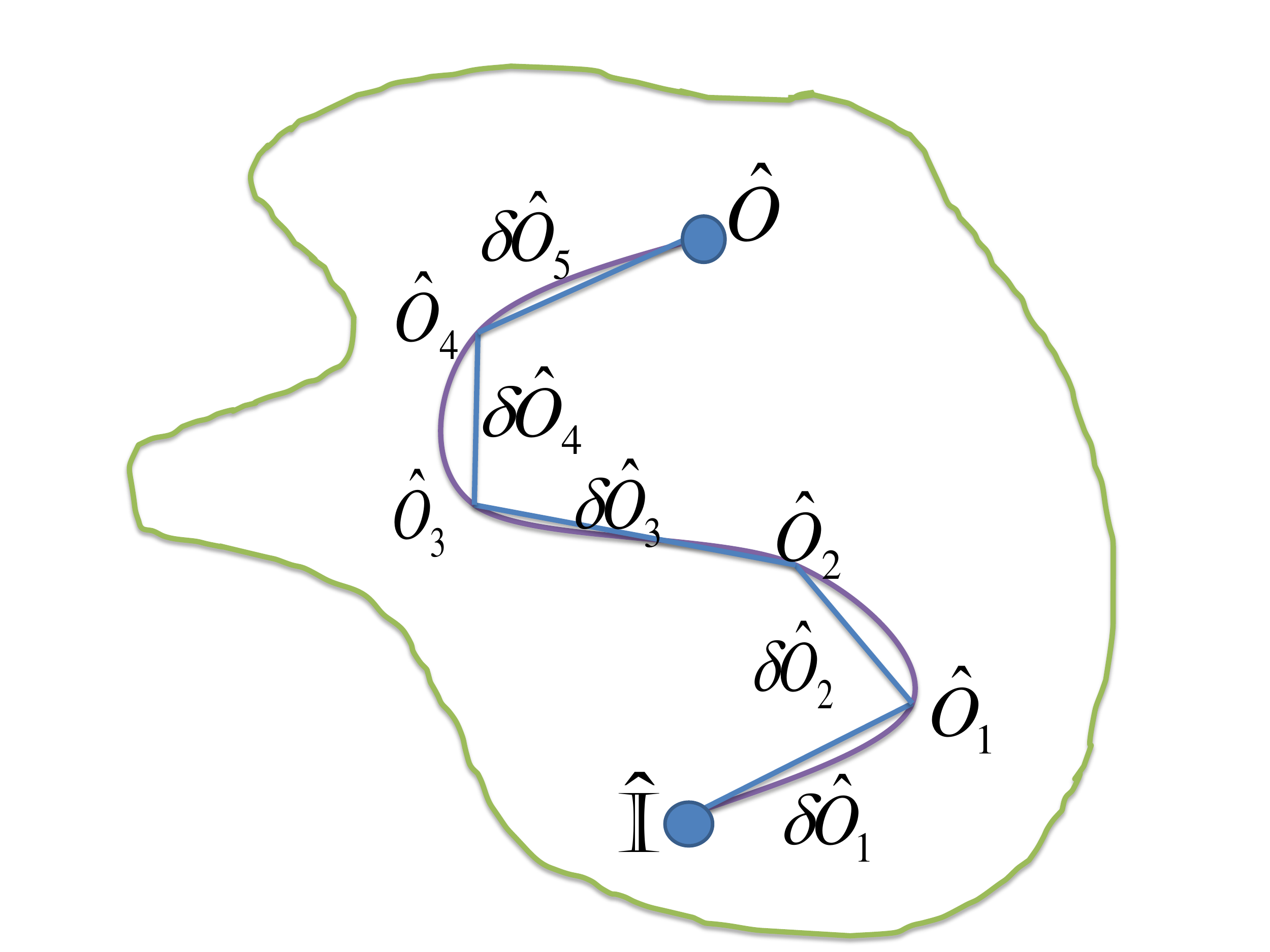}
  \caption{A continuous (purple) curve $c(s)$ connects the identity ($c(0)=\I$) and a particular operator $\hat{O}$ ($c(1)=\hat{O}$). It can be approximated by a discrete form (blue lines), where every intermediate point ($\hat{O}_n$) is also an operator.
} \label{expalinF1}
\end{figure}

\textbf{G4} stats that the complexity of an infinitesimal operator   is determined by its generator via some yet-unknown-function $\tilde{F}(H)$. Thus, one of important tasks is specifying the properties of $\tilde{F}(H)$ based on some physical requirements, which has been discussed in detail in \cite{Yang:2018nda,Yang:2018tpo}.
Once $\tilde{F}(H)$ is given, the complexity of ``finite'' operator can be computed, roughly speaking, by adding up the complexity of infinitesimal operator for the minimized path. To be more precise, let us consider the SU(n) manifold in Fig.~\ref{expalinF1}

\paragraph{Nielson's geometric idea for complexity}

As $\text{SU}(n)$ is connected, there is a curve $c(s)$ connecting $\hat{O}$ and identity $\I$, see Fig. \ref{expalinF1}, where the (purple) curve  is parameterized by $s$ with $c(0)=\I$ and $c(1)=\hat{O}$. Every intermediate point, $c(s) (0<s<1)$, also corresponds to an operator $\hat{O}(s)$ and we use the notation $c(s)$ and $\hat{O}(s)$ interchangeably depending on context. The discretized version of the curve is also shown as blue lines in Fig.~\ref{expalinF1}, where we use $c(s_n)$ or $\hat{O}_n$ interchangeably. Here, $s_n=n/N$, $n=1,2,3,\cdots, N$, $\hat{O}_{0}=\I$ and $\hat{O}_{N}=\hat{O}$.

For a given curve, it's discrete form can be represented in \textit{two} ways:
\begin{equation}\label{discreteO1}
\begin{split}
   c(s_n) =  \hat{O}_n  &=  \delta\hat{O}_n^{(r)}\hat{O}_{n-1} \\
    & =  \hat{O}_{n-1} \delta\hat{O}_n^{(l)}\,, 
\end{split}
\end{equation}
where
\begin{equation}
\delta\hat{O}_{n}^{(\alpha)}=\exp[-i H_\alpha(s_{n})\delta s] \,,
\end{equation}
with $\alpha=$ $r$ or $l$ and $\delta s=1/N$.
The continuous version of \eqref{discreteO1} is\footnote{It is important to note that left order and right order have nothing to do with changing time-ordering. The increasing direction of the parameter $s$, the time flow if you want, is the {\it same}, but the direction to multiply a new operator is different as shown in \eqref{discreteO1}. \label{okmj}}
\begin{equation}\label{defcs2}
  c(s)\ = \ \overleftarrow{\mathcal{P}}\exp\int_0^s-iH_r(\tilde{s})\td \tilde{s}\ = \ \overrightarrow{\mathcal{P}}\exp\int_0^s-iH_l(\tilde{s})\td \tilde{s}\,,
\end{equation}
%
In a diffential form, we have
\begin{equation}\label{rightH1}
\dot{c}(s) = -i H_r(s)c(s)  \,, \qquad \dot{c}(s) = -i c(s)H_l(s)  \,,
\end{equation}
or
\begin{equation}\label{rightH2}
H_r(s) = i \dot{c}(s) c(s)^{-1}\,, \qquad  H_l(s) = i c(s)^{-1}  \dot{c}(s)  \,,
\end{equation}
from which we obtain the relation between $H_r$ and $H_l$:
\begin{equation}\label{HrHl}
H_l(s)=c(s)^{\dagger}H_r(s)c(s)\,.
\end{equation}
Note that, in general, the two generators $H_r(s)$ and $H_l(s)$ at the same point of the same curve can be different.\footnote{One may think that the Schr\"{o}dinger's equation implies that only $H_r(s)$ has physical meaning and we have to use order $\overleftarrow{P}$. However, in Sec.~\ref{qonr} we will show that $H_l(s)$ is the Hamiltonian in the  Heisenberg picture so $H_l(s)$ and order $\overrightarrow{P}$ are also physically meaningful.}

%
%
%
%

We now want to compute the length of the curve ($L[c]$) from $c(0)$ to $c(1)$, of which meaning is the "cost" of the operator $\hat{O}$:  how much difficult it is to construct the operator $\hat{O}$. Once we know the complexities of the infinitesimal operators along the curve, we may sum them up. However, for a given curve, we may have two different costs(lengths) depending on the way to construct the target operator $\hat{O}$, ``left'' way or ``right'' wary, as shown in Eq.~\eqref{discreteO1} or Eq.~\eqref{defcs2}.
\begin{equation}\label{defLrLl1a}
  L_\alpha[c] :=\sum_{n=1}^N\C(\delta\hat{O}_{n}^{(\alpha)})=\sum_{n=1}^N\F(H_\alpha(s_{n}))\delta s\,, \qquad \alpha=r,l\,.
\end{equation}
In the continuous limit of $N\rightarrow\infty$, we have
\begin{equation}\label{defLrLl1b}
  L_\alpha[c]=\int_0^1\F(H_\alpha(s))\td s\,.
\end{equation}
Finally, the complexity is defined as the minimum length (cost) of the operator, i.e.
\begin{equation}\label{defCrl1}
  \C_\alpha(\hat{O}):=\min\{L_\alpha[c]~|~c(0)=\I,~~~c(1)=\hat{O}\}\,.
\end{equation}
Note that, two complexities ($\C_r$ and $\C_l$) may be very different for the same operator because $H_r\neq H_l$ in general. Let us call $\C_r$ right complexity and  $\C_l$ left complexity.

In Nielsen's geometrization of the complexity, only the right complexity is considered. All the other works such as Refs.~\cite{Brown:2017jil,Jefferson:2017sdb,Yang:2017nfn,Chapman:2017rqy,Khan:2018rzm,Camargo:2018eof,Chapman:2018hou,Balasubramanian:2019wgd} also considered only the right complexity. However, it will be a natural and interesting question what if we use the left complexity instead of right complexity. We will show they can be the same in some cases.

\paragraph{Constraint on the complexity metric and the complexity metric of SU(n)}

In order to perform concreted computations we need to know what $\tilde{F}(H_\alpha)$ is. As a first step, it has been shown~\cite{Yang:2018nda,Yang:2018tpo}, from the axioms \textbf{G1}, \textbf{G2}, and \textbf{G4}, that $\tilde{F}(H_\alpha)$ corresponds to a (Minkowski) norm\footnote{More precisely, it is called a \textit{Minkowski norm} in mathematical jargon. The properties that Minkowski norm should satisfy can be found in~\cite{038798948X,9810245319,xiaohuan2006an,asanov1985finsler} or below Eq.(3.7) in \cite{Yang:2018nda}.} of $H_\alpha$ in the Lie algebra.  It plays a role of the length of the line element in the Finsler geometry\footnote{{
For a given $\F$, we have two different natural ways to extend the Minkowski norm $\tilde{F}$ at the identity to every point on the base manifold via arbitrary curves.
\begin{equation} \label{rlinvar}
F(c, \dot{c})= \tilde{F} (H_r)=\F(\dot{c} c^{-1})\,, \  \ \mathrm{or} \ \ F(c, \dot{c})=\tilde{F} (H_l)=\F(c^{-1}\dot{c}) \,.
\end{equation}
where we introduce a notation `$F(c, \dot{c})$', the standard notation for {\it Finsler metric} in mathematics. We want to emphasize two points: (1) The Finsler metric is defined in the tangent bundle $TG$ for a Lie group $G$; (2) the norm $\tilde{F}(H)$ is defined at the tangent space of identity (i.e., the Lie algebra). For a Lie group and a given Minkowski norm $\tilde{F}(H)$, there are always two natural ways to obtain two different Finsler metrics according to Eq.~\eqref{rlinvar}. Understanding this point will be helpful to understand why bi-invariance is natural and may be necessary in physics. }}, which is a generalization of Riemannian geometry without the quadratic restriction.  We refer to Refs.~\cite{038798948X,9810245319,xiaohuan2006an,asanov1985finsler} for an introduction to  Minkowski norm and the Finsler geometry.

However, at this stage, any form of the Minkowski norm $\tilde{F}$ is allowed, so it will be desirable to have more physical conditions to constrain the form of $\tilde{F}$.  In Refs.~\cite{Yang:2018nda,Yang:2018tpo}, it has been further assumed that the norm $\F$ is invariant under a unitary transformation
\begin{equation}\label{unitarysy}
  \F(H_\alpha)=\F(\hat{U} H_\alpha\hat{U}^\dagger)\,,
\end{equation}
and reversal
\begin{equation}\label{cptsy}
  \F(H_\alpha)=\F(-H_\alpha)\,.
\end{equation}
The rationale for these assumption is that the complexity has to satisfy a general property of quantum mechanics/field theory and has to be consistent with general frameworks of quantum mechanics/field theory.
We refer to Refs.~\cite{Yang:2018nda,Yang:2018tpo} for many different arguments supporting \eqref{unitarysy} and \eqref{cptsy}. For example, we introduce a simple one based on the observation that physics is independent of ``pictures-(Schr\"{o}dinger or Heisenberg)'' in section \ref{qonr}.\footnote{The purpose of this paper is  not to justify Eq.~\eqref{unitarysy} and Eq.~\eqref{cptsy}. It has been reported in Refs.~\cite{Yang:2018nda,Yang:2018tpo} by several arguments. Our purpose here is to investigate the chaotic behavior of the complexity under the conditions Eq.~\eqref{unitarysy} and Eq.~\eqref{cptsy}.  For this purpose, they may be considered just conditions restricting the system we are considering.}

The assumptions \eqref{unitarysy} and \eqref{cptsy}, together with the axioms \textbf{G1}-\textbf{G4}, turn out to be strong enough to
determine the norm $\tilde{F}$ uniquely (up to an overall constant) as~\cite{Yang:2018nda,Yang:2018tpo}
\begin{equation}\label{biFinsM1}
  \F(H(s))=\lambda\left\{\Tr[(H(s)H^\dagger(s))^{p/2}]\right\}^{1/p},~~H(s):=H_l(s)~\text{or}~H_r(s)\,.
\end{equation}
Here, $c(s)$ is an arbitrary curve in SU($n$) group and $\lambda$ is an arbitrary positive constant. Note that $H_l$ and $H_r$ give the same complexity in this case. If $p=1$, the function $\F$ is just the ``trace norm''. If $p=2$, the function $\F$ gives the ``standard metric'' of SU(n) groups.

\paragraph{Invariance properties}
Let us turn to some invariance properties.
$\tilde{F}(H_r(s))$ is right-invariant, which means that $\tilde{F}(H_r(s))$ is invariant under the right-translation, $c(s)\rightarrow c(s)\x$ for all $\x\in$SU($n$). It is because $H_{r}$ is right invariant as shown in Eq.~\eqref{rightH2}.
Similarly, $\tilde{F}(H_l(s))$ is left-invariant.
However, if we require the unitary invariance \eqref{unitarysy}  we obtain
\begin{equation} \label{ppp1}
\tilde{F}(H_r(s)) = \tilde{F}(H_l(s)) \,,
\end{equation}
where we used Eq.~\eqref{HrHl}.
It means that we do not need to worry about two definitions, left or right complexity~\eqref{defCrl1}, because they are the same. It also means that $\tilde{F}(H_r)$ is both left-invariant and right-invariant. This property is called bi-invariant.

\paragraph{Complexity of SU(n) operator}
Thanks to the bi-invariance, the minimization in Eq.~\eqref{defCrl1} can be simplified dramatically.  It has been shown that, if the metric is  bi-invariance, the curve $c(s)$ is a geodesic if and only if there is a  constant generator $H(s) = \bar{H}$ such that~\cite{Latifi2013,Latifi2011}
\begin{equation}\label{geodesic1}
\dot{c}(s)=-i\bar{H}c(s) \ \  \text{or} \  \ c(s)=\exp(-is\bar{H}) \,.
\end{equation}
With the condition $\hat{O}=c(1)=\exp(-i\bar{H})$, we can solve $\bar{H}$  and so the complexity of $\hat{O}$ is given by
\begin{equation}\label{compforO}
  \mathcal{C}(\hat{O}) 
  = \min\{ \F(\bar{H})\ | \ \forall \, \bar{H}, s.t., \exp(-i\bar{H})=\hat{O}\}\,,
\end{equation}
where $\F$ is defined in Eq.~\eqref{biFinsM1}. The minimization `min' in  \eqref{compforO} in the sense of `geodesic' is already taken care of in  \eqref{geodesic1}.  Here, `min' means the minimal value due to non-uniqueness of the solution for the equation~$\exp(-i\bar{H})=\O$. As a concrete example for the minimization due to non-uniqueness, see Eq.~{\eqref{ytr}}.

For example, let us consider the SU(2) group in the fundamental representation. An operator $\hat{O}\in$SU(2) can be written as
\begin{equation}\label{relnaO}
  \hat{O} =  \exp(-i \theta\vec{n}\cdot\vec{\sigma}) =\I\cos \theta-i(\vec{n}\cdot\vec{\sigma})\sin \theta  \,, 
\end{equation}
where $\vec{n}$ is a unit vector, $\theta$ is a real number, and
$\vec{\sigma}:=(\sigma_x,\sigma_y,\sigma_z)$ are Pauli matrixes.
%
The values of $\bar{H}$ satisfying the condition in Eq.~\eqref{compforO} are multiple and we label them by the subscript $m$, $\bar{H}_m$:
\begin{equation}\label{solvesu2}
\bar{H}_m =   \left(\arccos[\text{Tr}(\hat{O})/2]+2m\pi\right)\vec{n}\cdot\vec{\sigma}  \,, 
\end{equation}
for $\forall m\in \mathbb{N}$.
The norm of $\bar{H}_m$, Eq.~\eqref{biFinsM1}, reads
\begin{equation} \label{ytr}
\tilde{F}(\bar{H}_m)  = \lambda\left\{\Tr[(\bar{H}_m \bar{H}_m^\dagger)^{p/2}]  \right\}^{1/p} = \lambda 2^{1/p}  \left(\arccos[\text{Tr}(\hat{O})/2]+2m\pi\right) \,,  
\end{equation}
The complexity of $\hat{O}$ is given by the minimum value of $\tilde{F}(\bar{H}_m)$ (Eq.~\eqref{compforO}):
\begin{equation}\label{compOsu2}
  \mathcal{C}(\hat{O})= \arccos[\text{Tr}(\hat{O})/2] \,,
\end{equation}
where we dropped the overall constant $\lambda 2^{1/p}$ since we can always define the complexity up to an overall constant.

\section{Time dependent complexity in the SYK model} \label{sec3}

It is sometimes claimed~\cite{Brown:2017jil,Balasubramanian:2019wgd} that the bi-invariant complexity can not show the linear growth of the complexity in an exponential time order. In the following subsection, we will show this may not be the case by providing a concrete counter example. It seems that the bi-invariant complexity is still a viable and competitive one in this respect.

The analysis on the SYK model in this subsection is inspired by a  recent very interesting paper~\cite{Balasubramanian:2019wgd}.
The SYK model is a quantum-mechanical system comprised of $N$ (an even integer) Majorana fermions $\chi_i$ with the Hamiltonian
\begin{equation}\label{defsykH1}
  H(\mathcal{J},N)=\sum_{i<j<k<l}^{N}J_{ijkl}\chi_i\chi_j\chi_k\chi_l\,,
\end{equation}
where the coefficients $J_{ijkl}$ are drawn at random from a Gaussian distribution with mean zero and variance $\sigma^2$
\begin{equation}\label{defsigamsyk}
  \sigma^2=\frac{6\mathcal{J}^2}{N^3}\,.
\end{equation}
Here $\mathcal{J}$ is a model parameter and describes the coupling strength. This model is expected to be chaotic and holographically dual to 2D quantum gravity~\cite{Kitaev,Maldacena:2016hyu,Polchinski:2016xgd,Kitaev:2017awl}. This model was used as a toy model to verify the complexity theory.

In the following discussions, we will focus on the case $p=1$ in axiom \textbf{G3}. We will make a comment on other choices of $p$ later. As discussed in the subsection, for a unitary operator $\U(t)=\exp(-iH(\mathcal{J},N)t)$, its complexity is given by
\begin{equation}\label{eqCU1}
  \C(t)=\min\left\{\lambda\left.\Tr\sqrt{VV^\dagger}~\right|~\forall V,~~s.~t.~\exp(-iV)=\U(t)=\exp(-iH(\mathcal{J},N)t)\right\}\,.
\end{equation}
Here we first take $\lambda=1$. Note that the complexity in Eq.~\eqref{eqCU1} is no longer the minimal geodesic of a Riemannian geometry but the minimal geodesic of a Finsler geometry. To compute the complexity, the first step is to solve all possible generators $V$. This can be done as follows. Suppose that  the eigenvalues and eigenstates of $H$ defined in Eq.~\eqref{defsykH1} to be $E_n$ and $|n\rangle$ with $n=1,2,\cdots 2^{N/2}$. Then we have
$$\U(t)=\sum_{n=1}^{2^{N/2}}e^{-iE_nt}|n\rangle\langle n|\,,$$
which is a diagonal form  so
\begin{equation}\label{operatorV1}
  \exp(-iV)=\U(t)\Rightarrow V=\sum_{n=1}^{2^{N/2}}(E_nt+2\pi k_n)|n\rangle\langle n|\,,
\end{equation}
with $k_n\in\mathbb{N}$. However, to keep $V\in\mathfrak{su}(2^{N/2})$, we need to have $V$ to be traceless and so there is a constraint on the integers $k_n$
\begin{equation}\label{constrk1}
  \sum_{n=1}^{2^{N/2}}k_n=0\,.
\end{equation}

As the operator $V$ in Eq.~\eqref{operatorV1} has a diagonal form, Eq.~\eqref{eqCU1} becomes
\begin{equation}\label{eqCU2}
  \C(t)=\min\left\{\left.\sum_{n=1}^{2^{N/2}}|E_nt+2\pi k_n|\right|~\forall k_n\in\mathbb{N},~~s.~t.~\sum_{n=1}^{2^{N/2}}k_n=0\right\}\,.
\end{equation}
The eigenvalues $E_n$ can be obtained numerically by the exact diagonalization of the Hamiltonian~\eqref{defsykH1} up to $N\sim32$. In the case $2^{N/2}\gg1$, the above minimization can be approximated by
\begin{equation}\label{eqCU2}
  \C(t)\approx\sum_{n=1}^{2^{N/2}}\left|E_nt-2\pi [[E_nt/(2\pi)]]\right|\,,
\end{equation}
where the notation $[[X]]$ stands for the most neighboring integer of $X$. For example, $[[1.2]]=1, [[1.7]]=2$ and $[[-2.7]]=-3$.\footnote{This approximation can be understood as follows. If we assume $k_n$ is not integer then $\C(t)$ will be minimized when $k_n = -E_nt/(2\pi)$. To make $k_n$ integers we introduce the operation $[[X]]$. In this case $\sum_n k_n$ may not vanish exactly. However, for large $n$, $\sum_n k_n \approx 0$ on average.}

\paragraph{Digression: fixing total energy} In most theoretical studies about the SYK model, one usually fix the parameter $\mathcal{J}$ and study how the system depends on the fermion number $N$. However, this may not be the case in studying the complexity. In the study of the complexity, the physical question we may ask is  ``For an isolated system driven by a given energy $E$, how fast can the complexity of the system change?''

Although we do not need to introduce a concept of ``total energy''  to define  the complexity geometry, we need to inject the energy to the system to drive it to evolve. For isolated systems, we only need to inject energy only at the initial time as the total energy will be conserved; in disscipated systems, we need to keep injecting energy.

For example in quantum circuits, the Hadamard gate $g_H$ is one of the fundamental gates, which transforms one qubit states $|0\rangle\rightarrow(|1\rangle+|0\rangle)/\sqrt{2}$ and $|1\rangle\rightarrow(|1\rangle-|0\rangle)/\sqrt{2}$. In one-qubit Hilbert space, its representation reads
\begin{equation}\label{defgH1}
  g_H=\frac1{\sqrt{2}}\left[\begin{matrix}
  1,&1\\
  1,&-1
  \end{matrix}
  \right] \,,  \quad \mathrm{with} \  |0\rangle=\left(\begin{matrix}
  1\\
  0
  \end{matrix}
  \right) \,, \
    |1\rangle=\left(\begin{matrix}
  0\\
  1
  \end{matrix}
  \right) \,.
\end{equation}
From mathematics perspective, it is simply a well-defined matrix. However, in physical situations, we have to use a quantum mechanical system to realize it. This means we have to create an interaction system with some Hamiltonian $V$ and stop the interaction after a time $t$ so that  $e^{iH't}=g_H$. We find that
\begin{equation}
H'=E_1|e_1\rangle\langle e_1|+E_2 |e_2\rangle\langle e_2| \,,
\end{equation}
where
\begin{equation}
|e_1\rangle= \frac{1}{\sqrt{4+2\sqrt{2}}}  \left(\begin{matrix}
  1+\sqrt{2}\\
  1
  \end{matrix}
  \right) \,, \quad
  |e_2\rangle= \frac{1}{\sqrt{4-2\sqrt{2}}}  \left(\begin{matrix}
  1-\sqrt{2}\\
  1
  \end{matrix}
  \right)  \,,
\end{equation}
are two eigenvectors of $g_H$ and\footnote{In principle,
\begin{equation}
E_1t=2k_1\pi,~~E_2t=\pi+2k_2\pi\,,
\end{equation}
{with  integers $k_1$ and $k_2$.} is allowed, but we do not consider it because we are interested in the first moment that the Hadamard gate is realized. }
\begin{equation}
E_1=0\,, \quad E_2t=\pi \,.
\end{equation}
As a result
%
%
%
%
\begin{equation}
t = \frac{\pi}{E_2} = \frac{\pi}{\Delta E}  \,,
\end{equation}
where $E:=|E_2-E_1|$.
We see that, the time to finish one gate operation (i.e. the complexity of the system then will increase by 1) is not unique and depends on the the value of $\Delta E$. From the physical viewpoint, $\Delta E$ is just the difference between the exited state and the ground state of the Hamiltonian $H'$, which is the energy that the system can absorb or emit. Without fixing this energy, it will be ambiguous to talk about the complexity growth rate.

In more complicated situations, such as quantum computations, we want to see how quickly the target state can be achieved after injecting energy $E$ into the system. To finish the same computational task, we can always use more energy to reduce the required time. Thus, without fixing the input energy, comparing computational times of different circuits is not well-defined. This is also the case for the complexity, i.e., Fixing the available energy will be important to compare the evolutional time.

From the mathematical viewpoint, this issue can be seen more clearly. For a given geodesic evolution, the complexity is just this geodesic length if we assume there is no conjugate point.  Though this geodesic (strictly speaking, it is the image of the geodesic in SU($n$) group) and its length are unique and do not depend on  parameterizations, the complexity growth rate depends on  parameterizations. For example, two parameterizations $c_1(s)=e^{-iHs}$ and $c_2(s)=e^{-i\gamma Hs}$ in fact stand for the same image in SU($n$) group but they have different complexity growth ``rates''. It is ambiguous to compute the complexity growth rates without fixing this freedom.

\paragraph{The setup for the SYK model}
Let us return to the SYK model.
To compute the ``injected'' energy into the system, we shift the ground state energy to zero. Thus, the shifted Hamiltonian is
\begin{equation}\label{realphH}
  \mathcal{H}(\mathcal{J},N)=H(\mathcal{J},N)-E_{\min}(\mathcal{J},N) \,,
\end{equation}
where $E_{\min}(\mathcal{J},N)$ is the ground state energy of $H(\mathcal{J},N)$.  The eigenvalues of the shifted Hamiltonian reads
\begin{equation}
\mathcal{E}_n=E_n-E_{\min}(\mathcal{J},N)\,.
\end{equation}
When we consider the operator complexity, since all the eigenstates equally take part in the evolution, the total energy involved in the systems is given by
\begin{equation}\label{eqforEN1}
\begin{split}
 \langle E \rangle=&\langle\sum_{n=1}^{2^{N/2}}\mathcal{E}_n\rangle=\langle\sum_{n=1}^{2^{N/2}}(E_n-E_{\min}(\mathcal{J},N))\rangle\\
  =&\langle\Tr(H(\mathcal{J},N))\rangle-2^{N/2}\langle{E}_{\min}(\mathcal{J},N)\rangle=-2^{N/2}\langle{E}_{\min}(\mathcal{J},N)\rangle\,,
  \end{split}
\end{equation}
where $\langle\cdot\rangle$ stands for the average value, since the SYK model contains the random coupling $J_{ijkl}$.
In short, we have the equation between he total available energy $ \langle E \rangle $,  the coupling $\mathcal{J}$, and the fermion number $N$
\begin{equation}\label{eqforJNE}
  \langle E \rangle = -2^{N/2}\langle E_{\min}(\mathcal{J},N)\rangle\,.
\end{equation}
In particular, it has been shown that, when $N\geq8$, the ground state energy of the Hamiltonian $\mathcal{H}(\mathcal{J},N)$ can be fitted well by the following linear relationship~\cite{Garcia-Garcia:2016mno}
\begin{equation}\label{linearEN1}
  \langle E_{\min}(\mathcal{J},N)\rangle \approx -(0.055+0.029N)\mathcal{J}\,,
\end{equation}
so we obtain
\begin{equation}\label{eqJNE}
  \langle E \rangle  \approx 2^{N/2}(0.055+0.029N)\mathcal{J}\,.
\end{equation}

\begin{figure}
  \centering
  \includegraphics[width=0.45\textwidth]{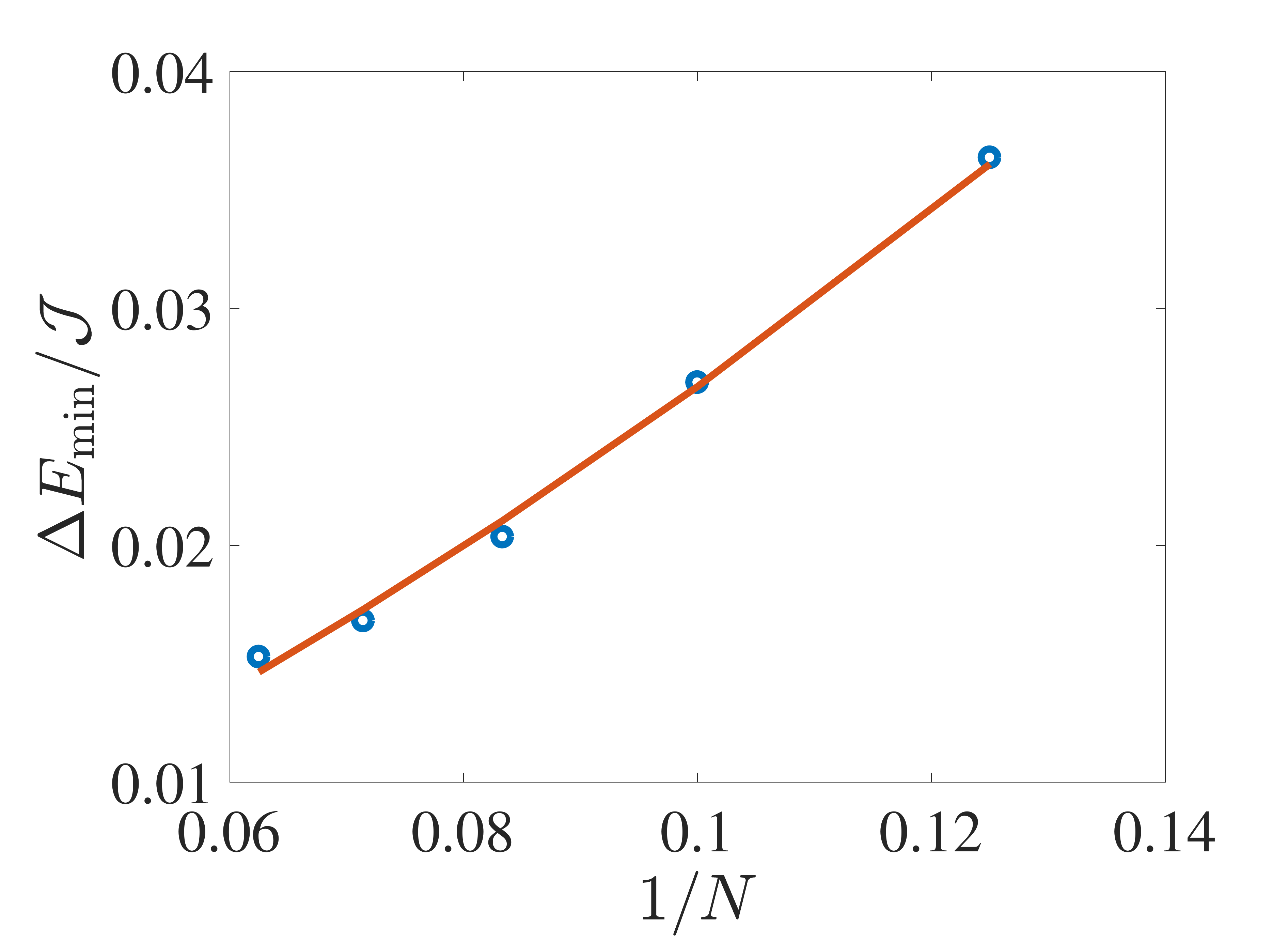}
  \caption{The relationship between $\Delta E_{\min}$ and $N$. The red line is the fitting curve $\Delta E_{\min}/\mathcal{J}=0.18N^{-1}+0.87N^{-2}$.  } \label{relDEN}
\end{figure}

To make the theory self-consistent, we need to insure that the fluctuation of the energy $\Delta E$ to satisfy $\Delta E/E\rightarrow0$ so that the system has a well defined energy. From Eq.~\eqref{eqforJNE} we see that
\begin{equation}\label{deltaE1}
  \Delta  E  =2^{N/2}\Delta E_{\min}\,,
\end{equation}
{where $\Delta E_{\min}:=\sqrt{\langle E_{\min}^2\rangle-\langle E_{\min}\rangle^2}$.} For large $N$, the relationship between $\Delta E$ and $N,\mathcal{J}$ can be obtained numerically, which is shown in Fig.~\ref{relDEN}. The fitting results show $\Delta E_{\min}\approx(0.18N^{-1}+0.87N^{-2})\mathcal{J}$ and so
\begin{equation}\label{relDENeq1}
  \Delta E\approx2^{N/2}(0.18N^{-1}+0.87N^{-2})\mathcal{J}\,.
\end{equation}
Taking Eq.~\eqref{eqJNE} into account, we find that, for large $N$, $\Delta E/\langle E \rangle \rightarrow\mathcal{O}(N^{-2})\rightarrow0$.

The Eqs. \eqref{eqJNE} and \eqref{relDENeq1} show there may be a subtle issue in the parameter fixing in Refs.~\cite{Magan:2018nmu,Brown:2017jil,Balasubramanian:2019wgd}, where  the coupling constant $\mathcal{J}$ is fixed and the large-$N$ limit is taken.  This means that the total available energy in the system and energy fluctuation both explode exponentially as $\mathcal{O}(2^{N/2})$.

\paragraph{Numerical method}
Note that the system has the following scaling symmetry
\begin{equation}\label{scaling}
  (\mathcal{J},\langle E \rangle ,t,\C)\rightarrow(\lambda\mathcal{J},\lambda \langle E \rangle ,\lambda^{-1}t,\C) \,.
\end{equation}
By this scaling transformation, we can first fix the coupling $\mathcal{J}$ to be unity in the numerical simulation and then transform the results into the case of fixing total energy $\langle E \rangle $. When we fix energy $ \langle E \rangle  $ to be of order $\mathcal{O}(1)$, the energy fluctuation $\Delta E$ is suppressed to be of order $\mathcal{O}(1/N^2)$ for large $N$.

For a given fermion number $N$ and total energy $\langle E \rangle$, the steps of computing the complexity evolution then as follows:\footnote{Our computational steps agree with \cite{Balasubramanian:2019wgd} up to (3) but differs in step (4).}
\begin{enumerate}
\item[(1)] Generate the Gaussian random coefficients $J_{ijkl}$ with $\mathcal{J}=1$\footnote{We fix  $\mathcal{J}=1$ instead of  $\langle E \rangle=1$ because of it is simple for numerics.} and write down the matrix element of Hamiltonian $H(\mathcal{J},N)$;
\item[(2)] Numerically diagonalize $H(\mathcal{J},N)$ and find its eigenvalues;
\item[(3)] Use Eq.~\eqref{eqCU2} to find the complexity $\C(t)$ at a given time $t$;
\item[(4)] Use the scaling transformation~\eqref{scaling} to convert the result into the case of fixing total energy $ \langle E \rangle$;
\item[(5)] Repeat steps (1)-(4) many times so that the average of $\C(t)$  converges.
\end{enumerate}

\begin{figure}
  \centering
  \includegraphics[width=0.45\textwidth]{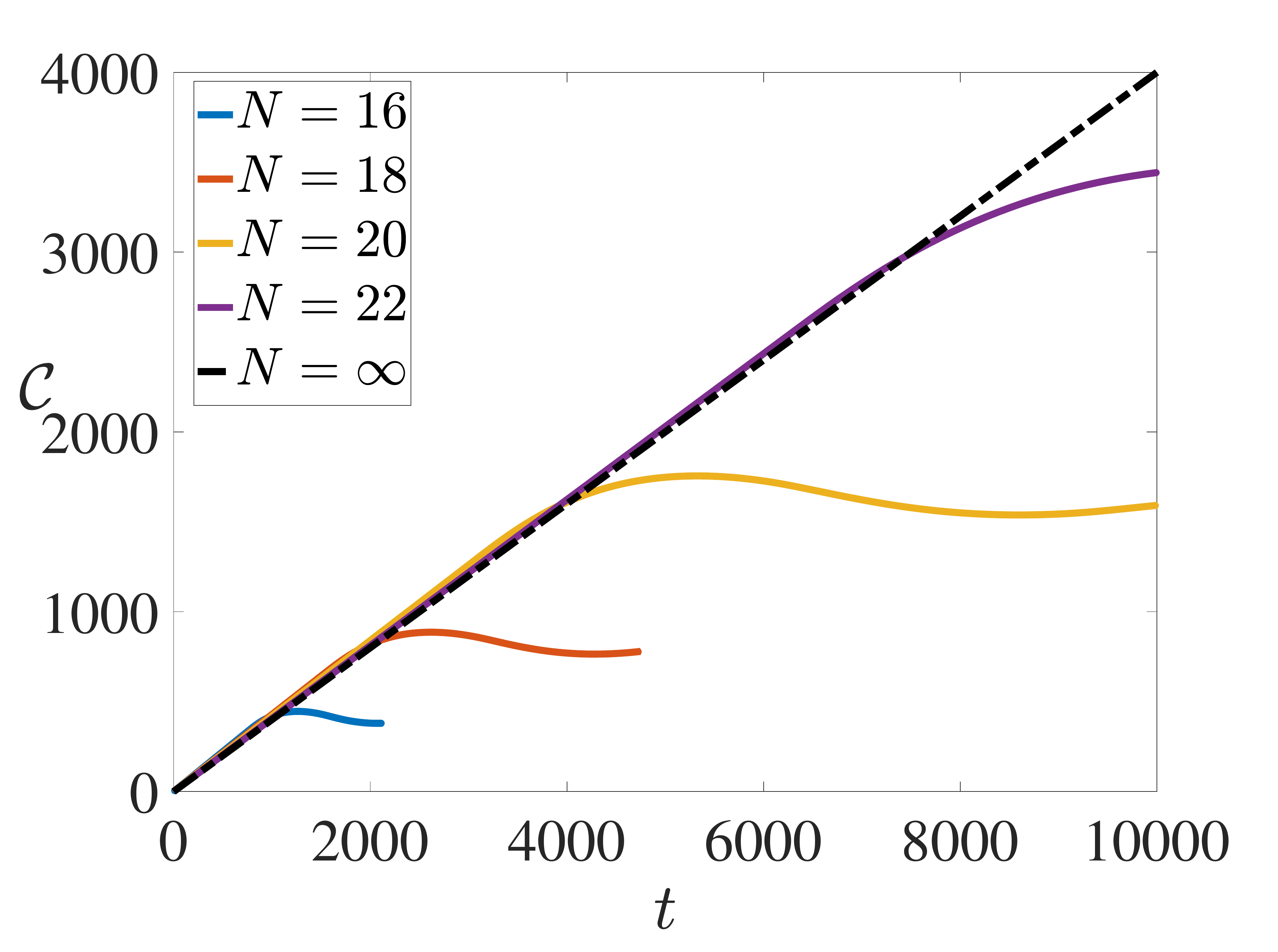}
  \includegraphics[width=.45\textwidth]{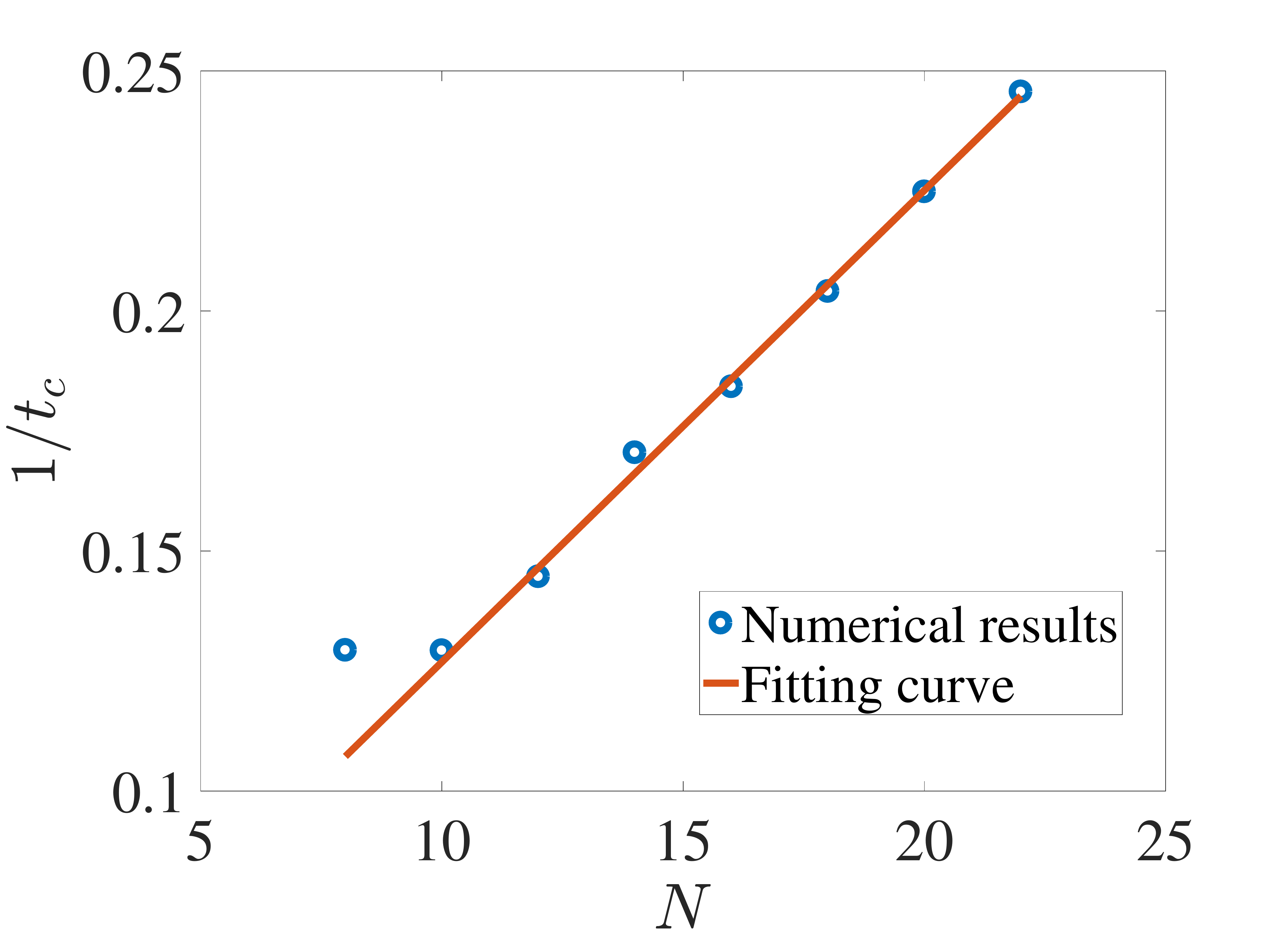}
  \caption{Left panel: the complexity growth when $N=16,18,20,22$ and $N\rightarrow\infty$ with $\langle E \rangle =1$. Right panel: the critical time $t_c$ vs the fermion number $N$ with a fixed $\mathcal{J}=1$. The red line is the fitting curve, which shows that $1/t_c|_{\mathcal{J}=1}\approx 0.01N+0.0265$.} \label{CttcN1}
\end{figure}

\paragraph{Linear growth and critical time} In the left panel of Fig.~\ref{CttcN1}, we show the complexity growth for $\langle E \rangle =1$ and $N=16,18,20, 22$. we find that the complexity grows linearly at early time. There is a critical time $t_c$ when the complexity stop growing linearly and go into a plateau with small fluctuation.

For a very large $N$ with fixing $\mathcal{J}=1$, it has been shown  that~\cite{Balasubramanian:2019wgd}, the linear growth time scale is of order $t_c\sim\mathcal{O}(1/N)$. The more exact relationship between $t_c|_{\mathcal{J}=1}$ and $N$ can be fitted by the ansatz $1/t_c|_{\mathcal{J}=1}=a+bN$. By using the numerical results of $N=10,12,14,16,18,20, 22$, we find (see right panel of Fig.~\ref{CttcN1})
\begin{equation}\label{reltcJ1}
  t_c|_{\mathcal{J}=1}\approx\frac{100}{2.65+N}\,,  \quad N\gg1\,.
\end{equation}
Transforming this into the case of fixing energy $\langle E \rangle =E_0$ according to \eqref{scaling}, we have
\begin{equation}\label{reltcJ2}
  t_c|_{\langle E \rangle =E_0}\approx\frac{2^{N/2}(5.5+2.9N)}{(2.65+N)E_0}\,, \quad N\gg1\,.
\end{equation}
In the large $N$ limit, we have
\begin{equation}\label{reltcN}
  t_c|_{\langle E \rangle =E_0}\approx\frac{2^{N/2}\times3}{E_0}\,, \quad N\gg1\,.
\end{equation}
As expected, the linear growth time has the exponential order with respective to $N$.

There is a simple way to understand both the linear growth and the critical time Eq.~\eqref{reltcN}. From Eq.~\eqref{eqCU2} we can see that, if
\begin{equation}
 t<\frac{\pi}{E_{\mathrm{max}}} \,, \quad \mathrm{where} \quad E_{\mathrm{max}} := \max|E_n| \,,
\end{equation}
then $[[E_nt/(2\pi)]]=0$. Thus, the complexity will grow linearly
\begin{equation}
\C(t)\approx\sum_{n=1}^{2^{N/2}}|E_n|t\,.
\end{equation}
This linear growth will be first interrupted when $[[E_nt/(2\pi)]]=1$, which corresponds to the time scale
\begin{equation}
t_c \sim \frac{\pi}{E_{\mathrm{max}}}\,. 
\end{equation}
For $t>t_c$, the smaller energy levels than $E_\mathrm{max}$ start contributing to $[[E_nt/(2\pi)]]$ more and more, which will cancel the increase by the term $\sum_{n=1}^{2^{N/2}}|E_n|t$. It makes the plateau and fluctuations for $\C(t)$.   All the above argument is for one event. After taking the average and $N \to \infty$, we have $\langle 1/E_\mathrm{max}  \rangle \rightarrow 1/\langle E_\mathrm{max}\rangle $, i.e.
\begin{equation} \label{999}
t_c \sim \frac{\pi}{\langle E_{\mathrm{max}} \rangle } \sim \frac{2^{N/2} \times \pi}{ \langle E \rangle } \,, 
\end{equation}
which is consistent with our numerical result, Eq. \eqref{reltcN}.

\begin{figure}
  \centering
  \includegraphics[width=0.45\textwidth]{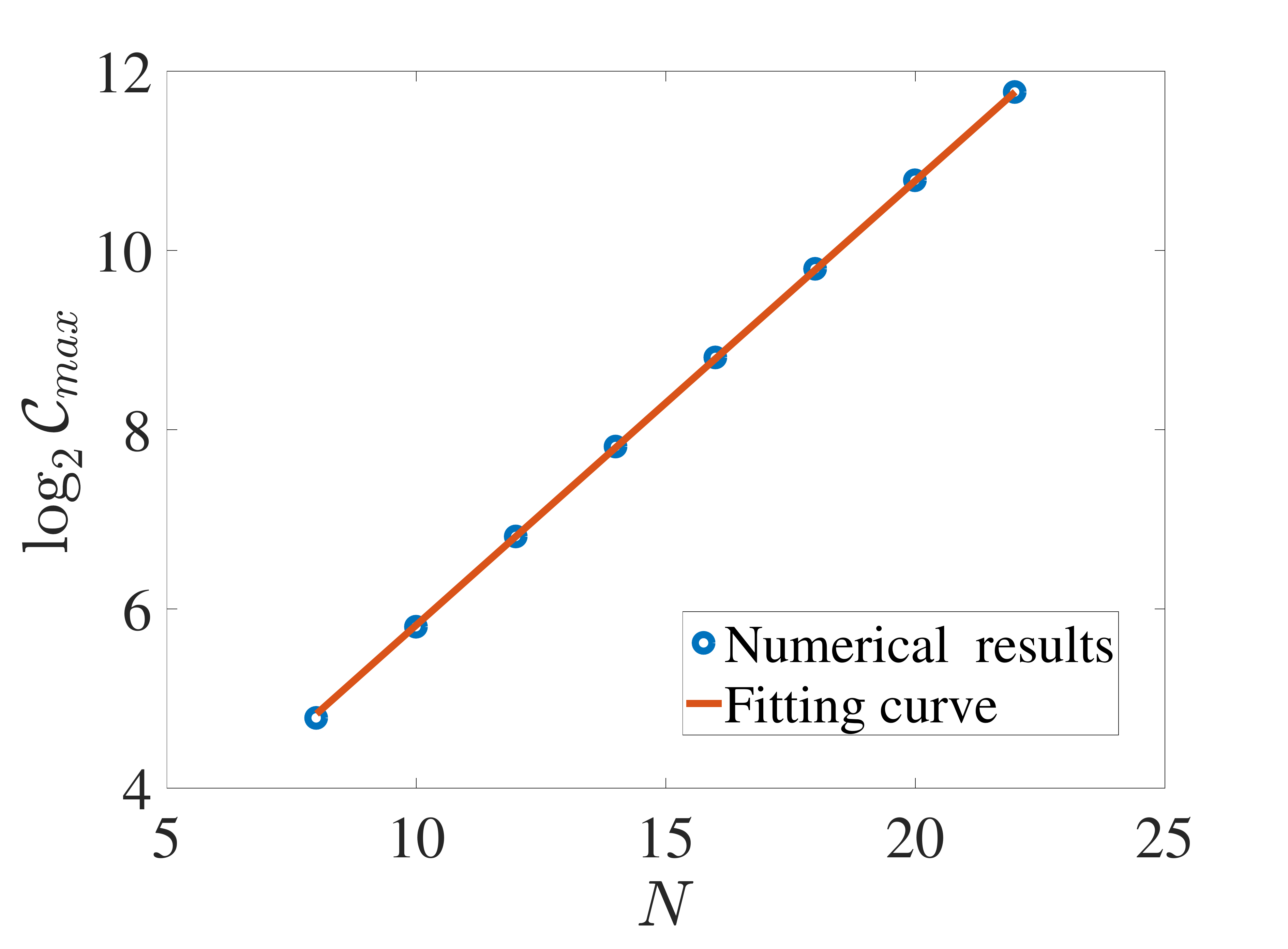}
  \includegraphics[width=0.45\textwidth]{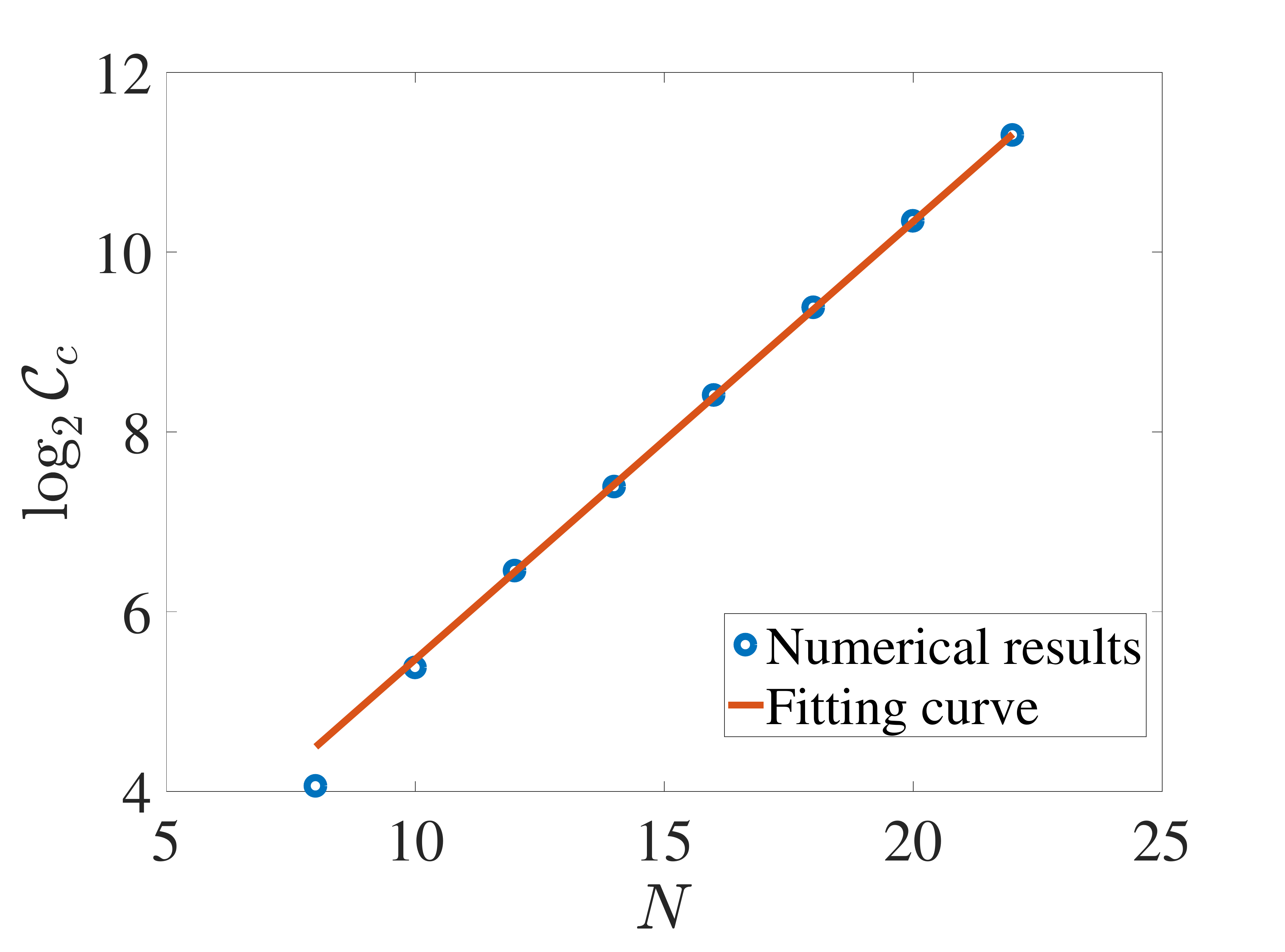}
  \caption{Left panel: relationship between the maximal complexity $\mathcal{C}_{\max}$ and the fermion number $N$. Right panel: relationship between the critical complexity $\mathcal{C}_{c}$ and the fermion number $N$.} \label{maxC1}
\end{figure}

\paragraph{Maximum or critical complexity and Lloyd's bound} We find that the maximum complexity $\C_{\max}$ and the critical complexity $\C_c$, which is defined by the complexity at the critical time $t_c$, also have the order of $2^{N/2}$. In  Fig.~\ref{maxC1}, we show the numerical results about the maximum complexity $\C_{\max}$ and the critical complexity $\C_c$, which give the following relationships
\begin{equation}\label{maxCeqN1}
  \C_{\max}(N)\approx1.82\times2^{N/2}\,, \quad \C_{c}(N)\approx1.5\times2^{N/2}\,.
\end{equation}
We also find that the complexity growth rate in the linear region in large $N$ limit reads
\begin{equation}\label{growthrate1}
  \left.\frac{\td\C(t)}{\td t}\right|_{\langle E \rangle =E_0}=\left.\frac{\C_c}{t_c}\right|_{\langle E \rangle =E_0}\approx0.5E_0\,, \quad N\gg1, ~~t<t_c.
\end{equation}
%
This says, for a given finite energy  in a large $N$ system, the complexity growth rate in its linear region is proportional to the total available energy. In addition, from our numerical results, we find the growth rate in its linear region is in fact the maximum growth rate.
By choose $\lambda\approx4/\pi$ in Eq.~\eqref{eqCU1}, we conclude that
\begin{equation}\label{growthrate1}
  \left.\frac{\td\C(t)}{\td t}\right|_{\langle E \rangle =E_0}\leq\frac{2E_0}{\pi}\,, \quad N\gg1 \,.
\end{equation}
This is nothing but the ``Lloyd's bound'', which was proposed by Ref.~\cite{Brown:2015bva,Brown:2015lvg} and was studied widely in holographic conjectures.\footnote{The recent studies~\cite{Carmi:2017jqz,Kim:2017qrq} show that, this bound is true only in the late time limit of the CA conjectures. At finite time, the holographic complexity in the CA conjecture can break this bound. To our knowledge, all reported results in the CV conjecture show that this bound is true for all time scale. {For one of the most recent studies see~\cite{Yang:2019gce} and see the references therein.}}

\paragraph{Why is $p=1$ favorable?} If we choose $p=2$ in the axiom \textbf{G3}, then the complexity geometry will be Riemannian geometry and we can do the similar analysis and find that $\C_{c}(N)\propto 2^{N/4}$ and $t_c(N)|_{\langle E \rangle =E_0}\propto2^{N/2}$. As a result, the complexity growth rate $\td\C(t)/\td t|_{\langle E \rangle =E_0}\rightarrow0$ when $N\rightarrow\infty$. In addition, for general $p$, we find that (see appendix~\ref{otherps} for an explanation)
\begin{equation}\label{Ccps}
  \C_{c}(N)\propto2^{N/(2p)}
\end{equation}
and $t_c(N)|_{\langle E \rangle =E_0}\propto2^{N/2}$ and
\begin{equation}\label{growthratep}
  \left.\frac{\td\C(t)}{\td t}\right|_{\langle E \rangle =E_0}=\left.\frac{\C_c}{t_c}\right|_{\langle E \rangle =E_0}\propto2^{\frac{N}{2p}-\frac{N}2}E_0\,, \quad N\gg1\,,  \quad t<t_c \,.
\end{equation}
Thus, only $p=1$ can give the constant finite nonzero complexity growth rate when we fix the energy $E$ and take the limit $N\rightarrow\infty$. This shows that the choice of $p=1$ (the most natural Riemannian Finsler geometry from the complexity perspective) is more favored than others (including mathematically simple Riemannian geometry) in describing the complexity of operators.

\paragraph{Quantum recurrence}
The Poincare recurrent theorem says: for any volume-preserving evolution in a compact manifold, the evolution will return to its arbitrarily small neighborhood at a finite time and the returning time increases exponentially by the number of degrees of freedom
\begin{equation}
t_r\sim\mathcal{O}(2^{2^{N/2}})\,.
\end{equation}
Any unitary evolution is a kind of volume-preserving evolution in a unitary group. When the unitary operator  approaches to the unit matrix, $\U(t_r)\approx\I$, so the complexity close to zero again. Due to the limitation of our numerical accuracy, we could not reach such a recurrence in our numerical analysis.


In summary, by a numerical simulation, we have shown that the bi-invariant complexity can yield two conjectures for the complexity growth of a chaotic system: i) linear growth until the exponential time scale, $e^N$; ii) after then, saturation and small fluctuation.
From the compactness of SU($n$) we may understand the quantum recurrence in a double exponential time scale, $e^{e^N}$.  We also confirmed the Lloyd's bound appears in our model.

It has been argued in Ref.~\cite{Balasubramanian:2019wgd} that the bi-invariant complexity does not exhibit the exponential time scale of linear growth. The main difference which leads us to obtain a different conclusion is that we ``fixed the total energy'' when we change the numbers of fermions, while in Ref.~\cite{Balasubramanian:2019wgd} the total energy is increased when  the numbers of fermions is increased.

The physical meaning of ``fixing total energy'' has been explained in more detail in the paragraph ``Digression: fixing total energy'' below Eq.\eqref{eqCU2}. Let us shortly rephrase it here. In the viewpoint of quantum computation, we can always obtain faster complexity growth by using more energy. For example, by using more energy we may use more computers simultaneously then we can do more computations per second (faster complexity growth).  Therefore, one good (or fair) way to compare the complexity growth rate of different systems is to ``fix the total energy''.

In the setup of Ref.~\cite{Balasubramanian:2019wgd},  more energy is injected to the system when the fermion number is increased. More injected energy yields faster complexity growth, which shortens the time scale of saturation, yielding polynomial time scale rather than the exponential time scale for the bi-invariant case. To overcome this issue,  Ref.~\cite{Balasubramanian:2019wgd} choose to deal with non-bi-invariant case while allowing ``increasing total energy''. However, another possibility is to consider ``fixing total energy'' with a bi-invariant case, which we have done in this paper.
In our setup, we fixed the available energy for SYK model, which is different from Ref.~\cite{Balasubramanian:2019wgd},  and showed that the exponential time scale can be indeed realized.


\section{Comment on bi-invariant complexity}\label{qonr}

In this section, we make some comments on the bi-invariance of the complexity. In fact, the bi-invariance (or non-bi-invariance) has something to do with the question we raise below Eq.~\eqref{defCrl1}, which is as follows.
\begin{itemize}
\item For a given curve, two {\it different} complexities (right-complexity and left-complexity) can be defined.
\item In most literatures, only the right-complexity is studied.
\item What if we consider the left-complexity? Is the physics equivalent to the right-complexity?
\end{itemize}
We think it is natural that the left complexity and right complexity are equivalent. If one believes that they are not equivalent, one needs to answer why the right-invariant quantity(length) corresponds to {\it the complexity} instead of left-invariant one. It seems that there is no a priori reason to prefer ``right'' to ``left''. As a related question, if one believes that the right-invariant length corresponds to {\it the complexity} it will be good to answer what the physical meaning of the left-invariant length is?

Because we do not have a good intuition for the case that the left and right are inequivalent, in this section, we restrict ourselves to the other case: the left-way and right-way must be equivalent. Now, the main question we want to ask is shifted to:
\begin{itemize}
\item[] If we believe the right invariant complexity and left invariant complexity are equivalent, what is the consequence of this equivalence regarding the complexity geometry?
\end{itemize}
Our answer is that the the complexity geometry needs to be unitary invariance and bi-invariance.

\paragraph{Why unitary invariance?}
First of all, we have shown that our formalism reviewed in section \ref{csyk} yields the equivalence of right-way and left-way so the bi-invariance of the complexity  in a natural way. In this result, the key point is the unitary invariance, Eq.~\eqref{unitarysy},  which implies the bi-invariance and the equivalence between the left and right quantities. See the discussion around Eq.~\eqref{ppp1}.

Here, we want to justify the unitary invariance a little more.
One may think it is more natural to use $H_r$ than $H_l$ because it is more {\it conventional} to use the first type of equation in Eq.~\eqref{defcs2} and Eq.~\eqref{rightH1}.
For example, the Schr\"{o}dinger's equation is conventionally written as
\begin{equation}\label{defschr01}
  \partial_t|\psi(t)\rangle=-iH_r(t)|\psi(t)\rangle\,,
\end{equation}
or, in terms of the time evolution operator, $\U(t)$ ($|\psi(t)\rangle \equiv \U(t)|\psi(0)\rangle$)
\begin{equation}\label{Uhs1}
\partial_t \U(t)=-iH_r(t) \U(t) \,,
\end{equation}
which is the same type as the first equation in Eq.~\eqref{rightH1}.

However, note that the equations~\eqref{defschr01} and \eqref{Uhs1} are the expressions in the Schr\"{o}dinger picture. Alternatively, we may use the Heisenberg picture without changing any physics. If $H_r(t)$ is the Hamiltonian in the Schr\"{o}dinger picture, the corresponding Hamiltonian in the Heisenberg picture ($H_H(t)$) reads
\begin{equation}\label{relHHHS}
  H_H(t)=\U(t)^\dagger H_r(t)\U(t) \,.
\end{equation}
Comparing it with Eq.~\eqref{HrHl} we find
\begin{equation} \label{nnn1}
H_H(t)=H_l(t)\,.
\end{equation}
Thus, the left-invariant Hamiltonian $H_l(t)$ is nothing but the Hamiltonian in the Heisenberg picture. In this sense $H_l(t)$ is as natural as $H_r(t)$. If we believe physics is independent of our choice of ``picture'', it is natural to expect that $H_r(t)$ and $H_l(t)$
give the same complexity.

Based on the observation from Eq.~\eqref{relHHHS} and Eq.~\eqref{nnn1} we can conclude that: if the  Hamiltonians in Heisenberg picture and  Schr\"{o}dinger picture are physically equivalent, then we have
\begin{equation}
\text{physics}(H(t_0)) = \text{physics}(U^\dagger(t_0) H(t_0) U(t_0)) \,,
\end{equation}
where $H(t_0)$ and $\U(t_0)$ stand for the Hamiltonian and evolutional operator at a given time $t=t_0$. Because $H(t_0)$ and $\U(t_0)$ are independent of each other at a given time $t=t_0$\footnote{Note that $\U(t_0)$ depends on the $H(t)$ with $t<t_0$ but is independent of $H(t_0)$. In other words, even though there is a relation between $H(t_0)$ and $\hat{U}(t_0)$, $H(t_0) = i \partial_t\hat{U}(t_0) \hat{U}(t_0)^{-1}$, $\partial_t\hat{U}(t_0)$ is arbitrary.} we conclude
\begin{equation} \label{poiu12}
\forall~H,~\forall~\U,~~~\text{physics}(H) = \text{physics}(U H U^\dagger) \,.
\end{equation}
Here, `physics' means physics studied so far except complexity theory.
Thus, we will assume the complexity theory also respect Eq.~\eqref{poiu12}, which means the unitary invariance of the norm  Eq.~\eqref{unitarysy}.\footnote{
See also ~\cite{Yang:2018nda,Yang:2018tpo} for several different supporting arguments for unitary invariance.} This is not a proof, but an argument to propose the unitary invariance assumption
Eq.~\eqref{unitarysy}.

\paragraph{Bi-invariance $\Leftrightarrow$ unitary invariance   $\Leftrightarrow$ left/right equivalence}

As a consequence of the unitary invariance, we have the equivalence of the right and left quantity (Eq.~\eqref{ppp1})
\begin{equation}
\tilde{F}(H_r) = \tilde{F}(H_l) \,,
\end{equation}
which also implies the bi-invariance of the length or complexity.
In fact, if we first assume the bi-invariance of the cost (curve length), it implies the unitary-invariance and the left/right equivalence of the norm $\F$. For example, for the right invariant norm, bi-invariance means
\begin{equation}
\tilde{F}(H_r) =  \tilde{F}( U H_r U^\dagger) \,,
\end{equation}
for arbitrary $U$. It is nothing but unitary invariance.
So far, we have shown the relation,
``bi-invariance $\Leftrightarrow$ unitary invariance  $\Rightarrow$ left/right equivalence''. To complete the equivalence relations we need to start with ``left/right equivalence'' to show unitary or bi-invariance. It has been proved in \cite{Yang:2018nda}.

\paragraph{More on left/right equivalence to unitary invariance or bi-invariance}
Note that so far we have used the same norm ($\tilde{F}$) for the left and right invariant case. This is the most natural choice in terms of the norms defined in a Lie algebra.
However, one may suspect that, if we allow the possibility of two different norms $\F_r$ and $\F_l$ for the right and left ways, the unitary invariance may not be necessary to have the equivalence between the left and right complexity because, in this case, naively it looks possible to have
\begin{equation} \label{uytrg}
\tilde{F}_r(H_r(s)) = \tilde{F}_l(H_l(s)) \,,
\end{equation}
so that $L_r[c] = L_l[c]$, without $\tilde{F}_r(H) = \tilde{F}_l(H)$ and without unitary invariance. See appendix~\ref{app111} for simple examples of non-unitary norms.

However, it can be shown that even in this case Eq.~\eqref{uytrg}, unitary invariance and bi-invariance are still implied i.e. we can prove the following theorem.
\begin{enumerate}
\item[] For a Lie group, suppose that we can define two different norms $\F_r$ and $\F_l$ for two generators $H_r$ and $H_l$ respectively, which give us two costs (lengths) $L_r[c]$ and $L_l[c]$ for a curve $c(s)$. 
If
\begin{equation}\label{convert1}
  \forall c(s)\,, \quad L_l[c]= L_r[c] \,,
\end{equation}
$\F_r$ and $\F_l$ must be unitary-invariant and so $L_r[c]$ and $L_l[c]$ are both bi-invariant.
\end{enumerate}
- proof step 1: let us first consider two curves $c_1(s)$ and $c_2(s)$. By Eq.~\eqref{convert1}, we have
\begin{equation}\label{convert2}
  L_l[c_1]=L_r[c_1]\,, \quad L_l[c_2]= L_r[c_2] \,.
\end{equation}
Taking the curve $c_2$ to be the right-translation of $c_1$, i.e., $c_2(s)=c_1(s)\U$, we have
\begin{equation}\label{convert3}
  L_l[c_1\U]= L_r[c_1\U] =  L_r[c_1] =L_l[c_1]\,,
\end{equation}
where in the second equality we used $L_r$ is right-invariant. This shows that $L_l$ is also right-invariant
\begin{equation}
L_l[c_1\U]=L_l[c_1] \,,
\end{equation}
and so bi-invariant. For the same reason, $L_r$ should be also bi-invariant.

\noindent - proof step 2: If $L_{l(r)}[c]$ is bi-invariant, it can be shown that  two norms $\F_{l(r)}$ is unitary invariant. 
To prove this, let us recall the definition of the right-invariant cost 
\begin{equation}
L_r[c]=\int_{0}^{1}\F_r(H_r(s))\td s\,, \quad H_r=i\dot{c}c^{-1}\,.
\end{equation}
If $L_r[c]$ is bi-invariant, it is invariant also under the left translation: $c(s)\rightarrow \U c(s)$, which means
\begin{equation}\label{relgb1}
  \int_{0}^{1}\F_{r}(H_r(s))\td s=\int_{0}^{1}\F_{r}(\U H_r(s)\U^{-1})\td s \,,
\end{equation}
for arbitrary curve $c(s)$. By choose $c(s)=\exp(-iH_{r0}s)$ with an arbitrary constant Hamiltonian $H_{r0}$, we find that Eq.~\eqref{relgb1} leads to
\begin{equation}\label{relgb2}
  \forall~H_{r0},~\forall~\U,~~\F_{r}(H_{r0})=\F_{r}(\U H_{r0}\U^{-1})\,.
\end{equation}
This proves that $\F_{r}(H_r)$ is unitary invariant. {By choosing $\hat{U} = c^{-1}$ in Eq.~\eqref{relgb2}, we also have bi-invariance of the norm $\F_{r}(H_r)$.}   Similarly, the proof works for $L_{l}[c]$ and $\F_{l}(H_l)$. Q.E.D.\footnote{Indeed, there are equivalence relations between unitary invariance and bi-invariance of a few quantities in complexity. We summarize and prove them in appendix \ref{bi-uni}. }

\begin{figure}
  \centering
  \includegraphics[width=.7\textwidth]{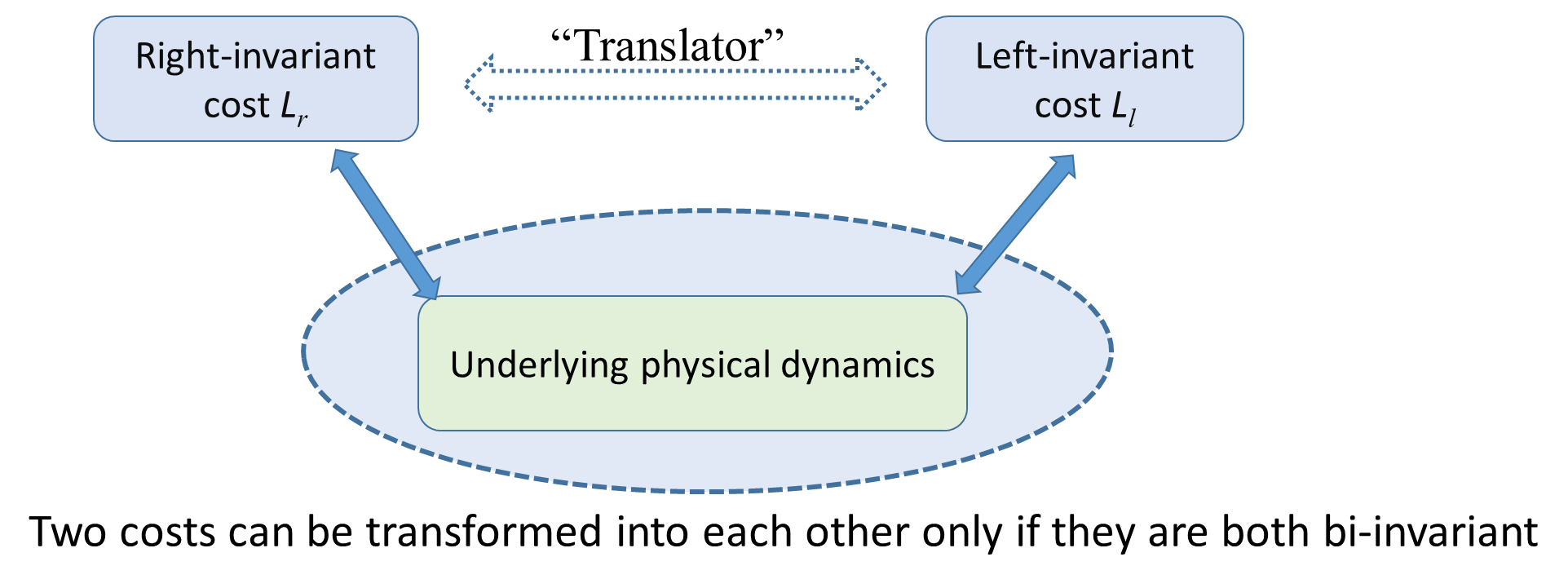}
  \caption{As two different costs describe the same underlying physical dynamics with a unique concept of complexity, there must be a ``translator'', which only depend on the complexity theory, not on individual systems. However, such a translator exists only if the complexity geometries ($\tilde{F}_{r(l)}$) are bi-invariant.} \label{trans1}
\end{figure}

One may further argue that even if $L_l[c] \ne L_r[c]$, it is still possible to describe the same physics if there is a relation
\begin{equation}
L_l[c]=\mathcal{T}(L_r[c])\,,
\end{equation}
where $\mathcal{T}(L_r[c])$ is a function of $L_r[c]$, which only depends on the complexity theory, not on individual systems. The function $\mathcal{T}(L_r[c])$ may be thought of as a ``translator'' between the two (left and right) complexity formalism describing the same underlying physics. see Fig.~\ref{trans1}.   For example, if $L_l[c] = 2 L_r[c] + 3$, it is still possible to extract the same information even though $L_l[c] \ne L_r[c]$. However, even in this case, the unitary invariance and bi-invariance are still necessary. The proof is the same as the one for Eq.~\eqref{convert1} with the replacement
$L_r[c] \rightarrow \mathcal{T}(L_r[c])$.

\paragraph{A consequence of non-bi-invariance}

As a corollary of the previous theorem \eqref{convert1} and its generalization $L_r[c] \rightarrow \mathcal{T}(L_r[c])$, we have:
\begin{itemize}
\item[] If the norm ($\tilde{F}_{r(l)}$) is not unitary-invariant or $L_{r(l)}[c]$ is not bi-invariant, we have
\begin{equation}\label{convert11}
   L_l[c] \ne L_r[c] \,,  \  \ \text{or more generally}  \,, \ \ L_l[c] \ne \mathcal{T}(L_r[c])\,,
\end{equation}
which means the left quantity and right quantity describe two different physics.
In this case, we have to face the question: which one corresponds to the length of the curve (cost of the operator) among left and right one and why?
\end{itemize}

We may rephrase the above statement in terms of complexity. If the norm $\tilde{F}$ (complexity geometry) is not unitary-invariant, the left complexity and the right complexity are not equivalent, so it seems not clear which one corresponds to {\it the complexity} in field theory. If we choose one (for what reason?), what will be the physical meaning of the other one? To our knowledge, these questions have not been asked seriously even though many people assume non-bi-invariance of complexity and study only right-invariant case.\footnote{One possible reason to prefer non-bi-invariant geometry has something to do with the concept of `locality'. In appendix \ref{commlocal}, we give some argument that the `locality' needs to be defined carefully.}
There may be good answers for these questions for non-bi-invariant cases. However, one possible resolution is to accept or assume the unitrary-invariance of the norm $\tilde{F}$ (and so the bi-invariance for the complexity geometry) as we have done in this paper, because it yields the equivalence of the left way and right way.

%
%

\section{Conclusions}
In this paper, we studied the time-dependent complexity of operators by the SYK model of $N$ Majorana fermions.  We have shown that the bi-invariant non-Riemannian Finsler geometry can be a good candidate for a complexity geometry in the sense that it exhibits three ``expected'' properties:
\begin{enumerate}
\item linear growth until the time scale, $t_c \sim e^N$
\item saturation and fluctuation after then
\item the quantum recurrence in a double exponential time scale, $t_r \sim e^{e^N}$
\end{enumerate}
Our results on the SYK model are inspired by the interesting research~\cite{Balasubramanian:2019wgd}, where a bi-invariant case for the SYK model was also discussed. (The previous research discussing the bi-invariant complexity in more general and detail are \cite{Yang:2018nda, Yang:2018tpo, Yang:2018cgx}.) Our numerical method basically agree with Ref.~\cite{Balasubramanian:2019wgd}, but the interpretation of the {\it complexity growth rate} is different. It is due to the difference of the reference time scale coming from the constraint: fixing total energy or fixing coupling of the SYK model. Note that the reference time scale is important because  the complexity growth rate is dimensionful, while the complexity is dimensionless and invariant under reparametrization.

We propose to fix the total energy instead of the coupling of the SYK model, because fixing energy is a natural physical condition to analyze the complexity. With this condition, we find that i) the critical time scale for the linear growth of the complexity is of order $e^N$; ii)  the complexity at this time scale is also of order $e^N$; iii) these two factors are canceled so the Lloyds' bound is achieved.

Furthermore, we have also shown that, as a complexity geometry,  the ``non-Riemannina'' Finsler geometry is more suitable than the widely used Riemannian geometry. Though in Nielsen's original work~\cite{Nielsen:2006:GAQ:2011686.2011688}, it has been already found that the Finsler (non-Riemannian) geometry is more suitable than the Riemannian geometry, most recent literatures including Nielsen himself's still used Riemannian geometry for simplicity. We showed that
only the most natural\footnote{The most natural means $p=1$ case in Eq.~\eqref{biFinsM1}.} (from the complexity perspective) ``non-Riemannian'' Finsler geometry can capture physics of the expected time evolution of the complexity.  This serves as the concrete first example showing the importance of the ``non-Riemannian''  Finsler geometry.

It is sometimes claimed that the bi-invariant complexity can not give the expected time-dependence of the complexity (see three items (1,2,3) above). However, in this paper, we have presented a counter example.
Thus, from this perspective, the bi-invariant complexity is still a  competitive candidate for the complexity in quantum mechanics and quantum filed theory. To give more intuitions on bi-invariant complexity, we have investigated an issue in defining the geodesic length in complexity geometry:  i) the complexity of a given operator can be defined in two independent ways geometrically, right-invariant way or left-invariant way; ii) then which one corresponds to the complexity? If the complexity is bi-invariant, two are equivalent so there is no issue any more, but if the complexity is not bi-invariant, two are not equivalent so we need to answer which one corresponds to the complexity.

For further arguments supporting the bi-invariant complexity in quantum mechanics/quantum field theory (QM/QFT), we refer to Refs.~\cite{Yang:2018nda, Yang:2018tpo, Yang:2018cgx}. The  rational behind these works are as follows. QM/QFT have their own constraints which are not shared by Nielson's quantum circuits, so some concepts of Nielson's may not be transferred to the complexity of QM/QFT. We may need to take general properties of QM/QFT into account more seriously. Under this consideration, bi-invariance may play an important role.

\acknowledgments
The authors would like to thank Yu-Sen An, Chao Niu, Cheng-Yong Zhang, Vijay Balasubramanian, Matthew DeCross, Arjun Kar, Onkar Parrikar for very helpful discussions.
The work of K.-Y. Kim was supported by Basic Science Research Program through the National Research Foundation of Korea(NRF) funded by the Ministry of Science, ICT $\&$ Future Planning(NRF- 2017R1A2B4004810) and GIST Research Institute(GRI) grant funded by the GIST in 2019.  We also would like to thank the APCTP(Asia-Pacific Center for Theoretical Physics) focus program,``Holography and geometry of quantum entanglement'' in Seoul, Korea for the hospitality during our visit, where part of this work was done.

\appendix
\section{{Explanation about Eq.~\eqref{Ccps}}}\label{otherps}
If we choose general $p$, Eq.~\eqref{eqCU2} should be replaced by
\begin{equation}\label{eqCU2b}
  \C(t)^p\approx\sum_{n=1}^{2^{N/2}}\left|E_nt-2\pi [[E_nt/(2\pi)]]\right|^p\,.
\end{equation}
The critical time $t_c$, i,e., the critical time of linear growth, is determined by Eq.~\eqref{999}  approximately in large $N$ limit and independent of $p$. Thus, we have
\begin{equation}\label{Ccp1}
  \C_c^p\approx t_c\sum_{n=1}^{2^{N/2}}|E_n|^p\,.
\end{equation}
The energy level obeys a Gaussian distribution. Assume $a$ to be the variance and define $x_n=E_n/a$. Then we have
\begin{equation}\label{sumEns1}
  \langle\sum_{n=1}^{2^{N/2}}|E_n|^p\rangle=a^p\langle\sum_{n=1}^{2^{N/2}}|x_n|^p\rangle=\frac{2^{N/2}a^p}{\sqrt{2\pi}}\int_0^\infty x^pe^{-x^2/2}\td x=\frac{2^{N/2}a^{p}}{2\sqrt{2\pi}}\Gamma\left(\frac{1+p}2\right)\,,
\end{equation}
where the notation $\langle\cdot\rangle$ stands for the average since the SKY model contains random coupling. As $t_c$ is independent of $p$, Eq.~\eqref{Ccp1} implies
\begin{equation}\label{Ccp-p1}
  {\C_c}^p= \C_c|_{p=1}a^{p-1}\Gamma\left(\frac{1+p}2\right)\propto 2^{N/2}\,,
\end{equation}
where we have used the fact in Eq.~\eqref{maxCeqN1} that $\C_c|_{p=1}\propto 2^{N/2}$ and
\begin{equation}\label{sumEns2}
  \langle\sum_{n=1}^{2^{N/2}}|E_n|^p\rangle=a^{p-1}\Gamma\left(\frac{1+p}2\right)\langle\sum_{n=1}^{2^{N/2}}|E_n|\rangle\,.
\end{equation}
Note in this discussion, we fix the total available energy $\langle E\rangle$ rather than fixing $\mathcal{J}$.  Eq.~\eqref{Ccp-p1} implies Eq.~\eqref{Ccps}

\section{A simple example for non-bi-invariant norm} \label{app111}
In this appendix, let us consider simple examples of non-bi-invariant (and non-unitary-invaraiant) norm and some consequences of left/right invariance.

Variants of Eq.~\eqref{biFinsM1}  can be used as non-unitary-invariant norms. For example, let us consider the case with $p=2$ and introduce  fixed matrices $M_r$ and $M_l$ under trace as
\begin{equation} \label{rinv1}
\tilde{F}_r(H_r) = \sqrt{\Tr H_r M_r H_r^\dagger}  \,,
\end{equation}
\begin{equation} \label{linv1}
\tilde{F}_l(H_l) = \sqrt{\Tr H_l M_l H_l^\dagger} \,.
\end{equation}
Eq.~\eqref{rinv1} is only right invariant but not left invariant, while Eq.~\eqref{linv1} is only left invariant but not right invariant.  Note that  i) this non-bi-invariance is due the matrices $M_r$ and $M_l$; ii) $M_r$ and $M_l$ are given by hand and not determined by the first principle in most literatures. This seems to give people a ``hope'' that we can always choose $M_l$ suitably so that
\begin{equation}\label{Mlr1}
\tilde{F}_r(H_r)  = \tilde{F}_l(H_l) \,,
\end{equation}
so we do not need to worry about the issue in choosing left-invariance and right-invariance.

However, we will show that Eq.~\eqref{Mlr1} is true only if $M_r\propto M_l\propto\I$, which implies $\F_r$ and $\F_l$ are both unitary invariant so gives bi-invariant complexity. The proof is simple. Eq.~\eqref{linv1} can be written by Eq.~\eqref{HrHl} as
\begin{equation} \label{linv2}
\tilde{F}_l(H_l) = \sqrt{\Tr H_r U  M_l  U^\dagger H_r^\dagger} \,,
\end{equation}
where $U$ is arbitrary. Then Eq.~\eqref{Mlr1} implies
$$\Tr H_r U  M_l  U^\dagger H_r^\dagger=\Tr H_rH_r^\dagger$$
for arbitrary $U$ and $H_r$ and so we must have
\begin{equation}
M_l \propto \mathbb{I} \,.
\end{equation}
We have the same conclusion for $M_r$. This means that the norm must be unitary-invariant.
This is a simple example showing why bi-invariance appears naturally from left/right equivalence. One may argue that by using a different norm in Eq.~\eqref{linv1} it is still possible to use a non-bi-invariant norm keeping left/right equivalence. However, we showed that it is not possible in general in Eq.~\eqref{convert1}.

\section{Some equivalences between bi-invariance and unitary invariance}\label{bi-uni}
In this appendix, we discuss equivalences between bi-invariance and unitary invariance of a few quantities. Suppose that $L_l[\cdot]$ and $L_r[\cdot]$ are left-invariant and right-invariant lengths/costs in a Lie group, respectively. $\F_l(\cdot)$ and $\F_r(\cdot,)$ are two norms for left-invariant Hamiltonian $H_l$ and right-invariant Hamiltonian $H_r$, respectively. $\C_r(\cdot)$ and $\C_l(\cdot)$ are  left-invariant and right-invariant complexities, respectively.
We will prove the following propositions:
\begin{enumerate}
\item[(a)] If there is a ``translator'' $\mathcal{T}$ which transforms the left-invariant cost into right-invariant cost, i.e., $\forall c(s)$,  $\mathcal{T}(L_r[c])=L_l[c]$, then $L_l[\cdot]$ and $L_r[\cdot]$ are both bi-invariant;
\item[(b1)] If $L_{l(r)}[\cdot]$ is bi-invariant, then $\F_{l(r)}(\cdot)$ is unitary invariant {and bi-invariant};
\item[(b2)] If $\F_{l(r)}(\cdot)$ is unitary invariant, then $L_{l(r)}[\cdot]$ is bi-invariant;
\item[(c1)] If $\F_{l(r)}(\cdot)$ is unitary invariant, then $\C_{l(r)}(\cdot)$ is  unitary invariant;
\item[(c2)] If  $\C_{l(r)}(\cdot)$ is unitary invariant, then $\F_{l(r)}(\cdot)$ is  unitary invariant;
\item[(d)] If $L_r[\cdot]=L_l[\cdot]$, then $\C_r(\cdot)=\C_l(\cdot)$ and $\F_{r}(\cdot)=\F_{l}(\cdot)$.
\end{enumerate}

\paragraph{Proof:}
The proof for proposition (a) and (b1) have been given in section \ref{qonr} and proposition (d) is obvious. Let us give the proofs for propositions (b1), (b2), (c1) and (c2):

\paragraph{(b2)}~~If the norm $\F_{r}(\cdot)$ is unitary invariant,  Eq.~\eqref{relgb1} is true so $L_r[c]=L_r[\U c]$. This means that $L_r[\cdot]$ is also left-invariant so bi-invariant. This proof works for $L_l[c]$ too, so proposition (b2) follows.

\paragraph{(c1)}~~By definition, the complexity of $\U$ is
\begin{equation}
\C_r(U)=\min\int_{0}^{1}\F_{r}(H_r(s))\td s\,, \quad U=\overleftarrow{P}\exp\left(-i\int_0^1H_r(s)\td s\right)\,.
\end{equation}
After a unitary transformation $\U\rightarrow\hat{O}\U\hat{O}^{-1}$, we have
\begin{equation}
\C_r(\hat{O}\U\hat{O}^{-1})=\min\int_{0}^{1}\F_{r}(\tilde{H}_r(s))\td s\,, \quad \hat{O}\U\hat{O}^{-1}=\overleftarrow{P}\exp\left(-i\int_0^1\tilde{H}_r(s)\td s\right)\,.
\end{equation}
Because
\begin{equation}
\C_r(\hat{O}\U\hat{O}^{-1})=\min\int_{0}^{1}\F_{r}(\hat{O}H_r(s)\hat{O}^{-1})\td s\,,
\end{equation}
if the metric is unitary invariant, i.e.,Eq.~\eqref{relgb2}, we have
\begin{equation}\label{Cunitary}
  \forall~\U\,,  \forall~\hat{O},~~~\C_r(\hat{O}\U\hat{O}^{-1})=\C_r(\U)\,.
\end{equation}
This proves (c1).

\paragraph{(c2)}~~For an infinitesimal operator $\U=\exp(-iH_{r0}\varepsilon)$ with an arbitrary Hamiltonian $H_{r0}$, the complexity is given by
\begin{equation}
\C_r(U)=\F_{r}(H_{r0})\varepsilon+\mathcal{O}(\varepsilon^2)\,.
\end{equation}
This can be understood by the geometrical analogues: any infinitesimally short curve is a ``straight line''. Under an arbitrary unitary transformation $\U\rightarrow\hat{O}\U\hat{O}^{-1}=\exp(-i\hat{O}H_{r0}\hat{O}^{-1}\varepsilon)$, we still obtain an infinitesimal operator and the complexity reads,
\begin{equation}
\C_r(\hat{O}\U\hat{O}^{-1})=\F_{r}(\hat{O}H_{r0}\hat{O}^{-1})\varepsilon+\mathcal{O}(\varepsilon^2)\,.
\end{equation}
If the complexity is unitary invariant, then we have
\begin{equation}
\forall~H_{r0},~\forall~\hat{O},~~~\F_{r}(\hat{O}H_{r0}\hat{O}^{-1})=\F_{r}(H_{r0})\,.
\end{equation}
This proves  proposition (c2).\\


\section{Comment on locality: apparent locality vs intrinsic locality}\label{commlocal}
It has been argued in Ref.~\cite{Brown:2017jil} that the complexity has something to do with ``locality''. The concept of ``locality'' will be more clarified later, but for now, we note, roughly speaking, ``local'' theory is ``simple'' and ``non-local'' theory is ``complex''.\footnote{{In our opinion,  this local/simple and non-local/complex relation may not be so robust. In principle, it is possible to have ``less complex'' non-local operator than a simple operator. Therefore, ``more non-local'' and ``more complex'' may not have a strong relationship in general.}} However, the unitary transformation in general seems to change the ``locality'' of the theory so should change the complexity. Therefore, one may conclude the complexity is non-unitary invariant.

Based on this argument, many literatures have tried to deal with non-unitary invariant or non-bi-invariant complexity by choosing some parameters in their theory: for example, by choosing $M_r$ in Eq.~\eqref{rinv1} by hand. In this section, we want to show that
\begin{itemize}
\item There are two kinds of locality, the ``apparent locality'' and ``intrinsic locality" (we will present detailed definitions later). The apparent locality may vary under the unitary transformation but intrinsic locality will not.  
    We think the ``locality'' used in Ref.~\cite{Brown:2017jil}  is an ``apparent locality'.
\item  The apparent locality, though it is useful in some cases, it cannot grasp the essential differences between local theory and non-local theory {regarding the complexity.} For example, suppose that the ``apparently'' local theory $H_l$ becomes the ``apparently'' non-local theory $H_r$ by a unitary transformation \eqref{defcs2}. It means that the Hamiltonian in Schr\"{o}dinger picture and the Hamiltonian in Heisenberg may have different ``apparent locality''. In this case, how do we know if the evolution operator $c(s)$ in Eq.~\eqref{defcs2} stands for a local theory or non-local theory?  The logical answer to avoid contradiction in Eq.~\eqref{defcs2} will be that it corresponds to an ``intrinsic'' locality.
\end{itemize}
In this section we try to clarify two facts: (1) it can not be read simply from its mathematical formula whether a Hamiltonian (not a Hamiltonian density) describes a local theory or not~\footnote{Note the difference between the Hamiltonian density and Hamiltonian. Suppose that $O(x)$ is an operator defined in spacetimes point $x$. Then $O(x)$ can be a Hamiltonian density but cannot be a Hamiltonian. A Hamiltonian must be expressed in a way similar to $\int\td xO(x)$.}
and (2) ``intrinsical'' locality will never be changed under unitary transformation.


Let us now go into more details.
We start with clarifying the meaning of  ``locality''\footnote{The ``locality'' can have different meanings in other contexts.
First, in the context of the quantum states, it means that the corresponding wave functions is well localized, i.e., $\psi(x)\rightarrow0$ rapidly if $x\rightarrow\pm\infty$. In the context of the field operator $\phi(x)$ it has something to do with local commutativity or microscopic causality, i.e., two fields are space-like separated and the fields either commute or anticommute.} used in Ref.~\cite{Brown:2017jil}.
It means that the mathematical expressions of Hamiltonian or Lagrangian  contain only local interactions and finitely many derivatives. We will call it ``apparent locality'' or an ``apparently local theory''.  A theory will be called ``apparently non-local'' if it is not an apparently local theory.

For example, the following Lagrangian is apparently local
\begin{equation}
L_1=(\partial_t\phi(x,t))^2-W(\partial_x\phi(x,t))-V(\phi(x,t))\,,
\end{equation}
where $W$ and $V$ are arbitrary two smooth functions. The following three Lagrangians are apparently non-local
\begin{equation}\label{defLs2}
  L_{2}=(\partial_t\phi(x,t))^2-W(\partial_x\phi(x,t))-\phi(x,t)\phi(x+a,t)\,, \quad a\neq0\,,
\end{equation}
\begin{equation}\label{defLs3}
  L_3=(\partial_t\phi(x,t))^2-W(\partial_x\phi(x,t))-\int\td y\phi(y,t)\phi(x+y,t)\,,
\end{equation}
and
\begin{equation}\label{defLs4}
  L_4=(\partial_t\phi(x,t))^2-V(\phi(x,t))-\phi(x,t)\left(\sum_{n=0}^{\infty}\frac{a^n}{n!}\partial_x^n\phi(x,t)\right)\,, \quad a\neq0\,.
\end{equation}
The $L_4$ is apparently non-local theory as
$$\sum_{n=0}^{\infty}\frac{a^n}{n!}\partial_x^n\phi(x,t)=\phi(x+a,t)\,.$$

In general, we can also define ``apparent $k$-locality'' and an ``apparently $k$-local theory'', in which the Hamiltonian and Lagrangian contains interactions involving $k$ different points.
For example, $L_1$ is apparently 1-local, while $L_2, L_3$ and $L_4$ are all apparently 2-local.
The Sachdev-Ye-Kitaev model is a quantum-mechanical system comprised of $N$ (an even integer) Majorana fermions $\chi_i$ with the Hamiltonian
\begin{equation}\label{defsykH11}
  H_{\mathrm{SYK}}=\sum_{i<j<k<l}^{N}J_{ijkl}\chi_i\chi_j\chi_k\chi_l\,,
\end{equation}
where the coefficients $J_{ijkl}$ are drawn at random from a Gaussian distribution. This is apparently 4-local as it involves the interactions of four different points.

To explain why the apparent locality may not be intrinsic, let us consider a similar example in general relativity. We may ask if the following metric
\begin{equation}\label{metric1}
  \td s^2=-(1-e^{2x})\td t^2+2(te^{2x}-1)\td t\td x+(t^2e^{2x}-1)\td x^2 \,,
\end{equation}
describes a flat spacetime or not? Naively (or ``apparently'' in our terminology), the metric looks not flat because it is different from $\td s^2=-\td t^2+\td x^2$. However, after the following coordinates transformation
\begin{equation}
\tau=t+x,~~\xi=t e^{x}\,,
\end{equation}
the above metric becomes $\td s^2=-\td \tau^2+\td\xi^2$, which is indeed flat. As is well known, flatness cannot be easily understood simply by looking at the ``apparent'' form of metric components.

A similar reasoning may apply to ``locality."
Let us now ask if the following Lagrangian
\begin{equation}\label{defLeg1}
  L=\left[\int\td yh(x,y)\partial_t\phi(y,t)\right]^2-\left[\partial_x\int\td yh(x,y)\partial_y\phi(y,t)\right]^2-\left[\int\td yh(x,y)\phi(y,t)\right]^2 \,,
\end{equation}
is ``local'' or not.  Here the integration range is $-\infty<x<\infty$, the function $h(x,y)$ satisfies
\begin{equation}\label{prophxy}
  \partial_xh(x,y)=-\partial_yh(x,y),~~h(x,y)|_{x\rightarrow\pm\infty}=h(x,y)|_{y\rightarrow\pm\infty}=0\,.
\end{equation}
and there is a function $\tilde{h}(x,y)$ such that
\begin{equation}\label{prophxy2}
  \int\td xh(x,y_1)\tilde{h}(x,y_2)=\delta(y_1-y_2),~~\int\td xh(x_1,y)\tilde{h}(x_2,y)=\delta(x_1-x_2)\,.
\end{equation}
This theory is ``apparently non-local'' as it involves the interactions of different points. However, making a variable transformation
\begin{equation}\label{defpsiohi1}
  \psi(x,t)=\int h(x,y)\phi(y,t)\td y\,,
\end{equation}
and noting the fact
\begin{equation}
  \int\td yh(x,y)\partial_y\phi(y,t)=h(x,y)\phi(y,t)|_{y=-\infty}^{y=\infty}-\int\td y\partial_yh(x,y)\phi(y,t)=\int\td y\partial_xh(x,y)\phi(y,t)\,,
\end{equation}
we have
\begin{equation}\label{defLeg2}
\begin{split}
  L&=\left[\partial_t\int\td yh(x,y)\phi(y,t)\right]^2-\left\{\partial_x\int\td y[\partial_xh(x,y)]\phi(y,t)\right\}^2-\left[\int\td yh(x,y)\phi(y,t)\right]^2\\
  &=[\partial_t\psi(x,t)]^2-[\partial_x^2\psi(x,t)]^2-\psi(x,t)^2\,.
  \end{split}
\end{equation}
After a suitable variable transformation, we find that the new Lagrangian~\eqref{defLeg2} becomes ``apparently local''.

To be self-consistent, it is necessary to check that if the variable transformation can keep the canonical commutation (or anticommutation) relation or not. The canonical momentum of $\phi$ for the Lagrangian~\eqref{defLeg1} reads
\begin{equation}\label{defpiphi}
  \pi_\phi(x,t):=\frac{\delta L}{\delta\partial_t\phi(x,t)}=2h(x,y)\int\td zh(y,z)\partial_t\phi(z,t)\,.
\end{equation}
We see that the momentum depends on the value of $\partial_t\phi$ in the whole space. The quantization can be achieved by imposing the following canonical commutation (or anticommutation) relation
\begin{equation}\label{commutator1}
  [\phi(x_1,t),\pi_\phi(x_2,t)]=i\delta(x_1-x_2)\,.
\end{equation}
From the Lagrangian~\eqref{defLeg2} we can obtain canonical momentum of $\psi(x,t)$
\begin{equation}\label{defpipsi}
  \pi_\psi(x,t):=\frac{\partial L}{\partial\partial_t\psi(x,t)}=2\partial_t\psi(x,t)=2\int h(x,y)\partial_t\phi(y,t)\td y\,.
\end{equation}
Combining the orthogonal relationship~\eqref{prophxy2} and the relationship~\eqref{defpiphi}, we obtain
\begin{equation}
\pi_\psi(x,t)=\int\tilde{h}(x,y)\pi_{\phi}(y,t)\td y\,.
\end{equation}
We see that, under the variable transformation~\eqref{defpsiohi1}, the canonical momentum is transformed as
$$\pi_\phi\rightarrow\pi_{\psi}=\int\tilde{h}(x,y)\pi_{\phi}(y,t)\td y\,.$$
Then we can check the new variables $\psi$ and $\pi_\psi$ satisfy the same canonical commutation (or anticommutation) relation
\begin{equation}\label{commutator2}
\begin{split}
  [\psi(x_1,t),\pi_\psi(x_2,t)]&=\left[\int h(x_1,y_1)\phi(y_1,t)\td y_1,\int \tilde{h}(x_2,y_2)\pi_{\phi}(y_2,t)\td y_2\right]\\
  &=\int h(x_1,y_1)\tilde{h}(x_2,y_2)\td y_1\td y_2[\phi(y_1,t),\pi_{\phi}(y_2,t)]\\
  &=i\int h(x_1,y_1)\tilde{h}(x_2,y_2)\td y_1\td y_2\delta(y_1-y_2)\\
  &=i\int h(x_1,y_1)\tilde{h}(x_2,y_1)\td y_1=i\delta(x_1-x_2)\,.
  \end{split}
\end{equation}
Checking such a self-consistence is necessary as not all variable transformations keep the canonical commutation (or anticommutation) relation. If a variable transformation changes these canonical relations, it will change physics.

We have found that, by a suitable variable transformation, an apparently non-local theory~\eqref{defLeg1} can be changed into an apparently local theory~\eqref{defLeg2}. One may argue that, though in term of $\psi(x)$, the Lagrangian~\eqref{defLeg2} has a local form, the field $\psi(x)$ contains integration of $\phi(x)$ and Eq.~\eqref{defLeg2} should still be treated as a non-local theory. About this argument, we would like to point out that Eqs.~\eqref{defpsiohi1} and \eqref{prophxy2} imply
\begin{equation}\label{defpsiohi2}
  \phi(x,t)=\int \tilde{h}(x,y)\psi(y,t)\td y,~~\psi(x,t)=\int h(x,y)\phi(y,t)\td y \,.
\end{equation}
The field $\phi(x)$ is also the integration of field $\psi(x)$ so there is no  reason to say that only $\phi(x)$ can be treated as a physical field operator but $\psi(x)$ can not be. After we choose $\psi(x)$ as the field operator, the theory becomes apparently local. Or we can say that, the Lagrangian~\eqref{defLeg1} is apparently non-local because we choose a ``bad'' field operator.

Some apparently non-local theories can be transformed into apparently local theories by suitable variables transformations, but some apparently non-local theories can not. For example, the three Lagrangians defined in Eqs.~\eqref{defLs2}, \eqref{defLs3} and \eqref{defLs4} can not be written in terms of apparently local Lagrangians by variable transformations. This means that, though Lagrangian \eqref{defLeg1} and Langrangians \eqref{defLs2}~\eqref{defLs3}~\eqref{defLs4} are all apparently non-local, they have essential differences. On the other hand, all apparently local theories can be transformed into the apparently non-local theories by suitable variable transformations. There are some freedoms in choosing the field operators and making variables transformations, and the apparent locality depends on the choices of variables and variable transformations. 
The Lagrangian~\eqref{defLeg1} looks like non-local because we choose the ``bad'' field operator rather than the theory is really non-local, which is similar to the aforementioned metric example: the metric~\eqref{metric1} ``apparently'' (naively) looks like curved spacetime because we choose ``bad'' coordinates rather than the spacetime is really curved.

If we want the concept of locality to be defined by some intrinsic properties of physical theories it should be defined as a way which does not depends on any specific choice of field operator. It is similar to general relativity: the flatness should be defined by a manner which does not depend on any specific choice of coordinates. Thus, for a field theory, it is more useful to define an ``intrinsic locality'' in such way:
\begin{enumerate}
\item[]\textit{If there is one suitable variable transformation to transform a Lagrangian into an ``apparently local'' form keeping the canonical commutation (or anticommutation) relation, then the theory is intrinsically local; if such a variable transformation does not exist, then the theory is intrinsically non-local. }
\end{enumerate}
The intrinsic locality will not be changed by variable transformations.

The above definition gives us a way to verify if a theory is intrinsically local or not. However, it is difficult to verify the existence of such a variable transformation for a general complicated Lagrangian.
In general relativity, it is also difficult to verify if there is a coordinates transformation so that the metric components becomes the  Minkowski form. However, the Riemann tensor offers us a powerful tool to judge the flatness even if we do not know such coordinates transformation.
Do we have any method to verify the intrinsic locality for a  given arbitrary Lagrangian even if we do not know the corresponding variable transformation? This question seems very interesting in both mathematics and physics. We do not have a complete answer. However, here we would like to present a simple relevant proposition:
\begin{enumerate}
\item[]\textit{A given Lagrangian $L$ in term of a field operator $\phi$ describes an intrinsically local theory if and only if its generating functional $Z[J]$ equals to the generating functional of an apparently local theory.}
\end{enumerate}
It can be partly justified by the \textit{Wightman reconstruction theorem}.\footnote{Strictly speaking the Wightman reconstruction theorem is valid for free scalar and spinor theories.} 
Because $Z[J]$ is the same as the generating functional of an apparently local theory,  its all $n$-point functions are the same as the $n$-point functions of an apparently local theory. The \textit{Wightman reconstruction theorem} says that such two theories are different only up to a unitary transformation. This means that we can find a unitary transformation $\phi\rightarrow\psi=\hat{U}\phi\hat{U}^\dagger$, under which the Lagrangian $L$ becomes apparently local and  the canonical momentum transforms as $\pi_{\phi}\rightarrow\pi_{\psi}=\hat{U}\pi_{\phi}\hat{U}^\dagger$. Such a unitary transformation is just a linear transformation and keep the canonical commutation (or anticommutation) relation unchanged. Thus, the proposition follows. This proposition shows that the intrinsic locality is also encoded in the generating functional. As a direct corollary, we have a conclusion:
\begin{enumerate}
\item[]\textit{If a Hamiltonian $H$ describes an intrinsically local (non-local) theory, then its arbitrary unitary transformation $H\rightarrow\U H\U^\dagger$ still describes an intrinsically local (non-local) theory. }
\end{enumerate}
We see that, the intrinsic locality, like the unitary-invariant complexity, is the unitary invariant quantity of a theory. {The Schr\"{o}dinger Hamiltonian $H_r$ and Heisenberg Hamiltonian $H_l$ may have different apparent localities but always have same intrinsic locality! }

In general relativity, we know that the information of flatness is encoded in the Riemann curvature tensor. We have also found that the intrinsic locality is encoded in the generating functional. Then what is the essential property of the generating functional for an intrinsically local theory? We think this is an interesting question to be investigated more.

To conclude, we argue in this section that the locality discussed in many literatures such as Ref.~\cite{Brown:2017jil} may be a kind of ``apparent locality'', which depends on one's choice of field operator (or ``coordinate'') so may not be able to grasp the essential differences between the local theory and non-local theory. The locality should be defined in an intrinsic way. If the generating functional of a theory is the same as an apparently local theory, then the theory is intrinsically local. {Such an intrinsic locality is unitary invariant and is consistent with the unitary invariant complexity. }

\bibliographystyle{JHEP}
\bibliography{cforchoas}

\end{document}